\documentclass[11pt,showpacs,floatfix,superscriptaddress,nofootinbib,amssymb,amsmath,aps,prc,eqsecnum]{revtex4}
\usepackage[dvips]{graphicx}
\usepackage{latexsym,epsfig,bm,times,psfrag,subfigure}
\usepackage{color}
\usepackage{dcolumn}                 
\usepackage[bookmarksnumbered,bookmarksopen,colorlinks,citecolor=blue,linkcolor=blue]{hyperref} %

\renewcommand\r{\rho}

\renewcommand\t{\tau}

\newcommand\e{\epsilon}

\newcommand\m{\mu}

\newcommand\p{\pi}

\newcommand\s{\sigma}
\newcommand\f{\phi}

\newcommand\ve{\varepsilon}

\newcommand{\fig}[1]{Fig.~\ref{#1}}
\newcommand{\eq}[1]{Eq.~(\ref{#1})}
\newcommand{\eqs}[2]{Eqs.~(\ref{#1})-(\ref{#2})}

\newcommand\lb{\left(}
\newcommand\rb{\right)}
\newcommand\ls{\left[}
\newcommand\rs{\right]}

\newcommand{\lan}{\langle}
\newcommand{\ran}{\rangle}

\newcommand\pt{\partial}

\newcommand{\ie}{\emph{i.e.}}

\newcommand{\eg}{\emph{e.g.}}
\newcommand{\etc}{\emph{etc}}

\newcommand{\br}{{\mathbf r}}

\newcommand{\bv}{{\bf v}}

\newcommand{\bB}{{\bf B}}
\newcommand{\bE}{{\bf E}}
\newcommand{\bJ}{{\bf J}}

\begin{document}

\title{Event-by-event generation of electromagnetic fields in heavy-ion collisions}
\author{Wei-Tian Deng}\email{deng@fias.uni-frankfurt.de}
\affiliation{Frankfurt Institute for Advanced Studies, D-60438 Frankfurt am Main, Germany}
\author{Xu-Guang Huang}\email{xhuang@itp.uni-frankfurt.de}
\affiliation{Frankfurt Institute for Advanced Studies, D-60438 Frankfurt am Main, Germany}
\affiliation{Institut f\"ur Theoretische Physik, J. W. Goethe-Universit\"at, D-60438 Frankfurt am Main, Germany}

\date{\today}

\begin{abstract}
We compute the electromagnetic fields generated in heavy-ion collisions by using the HIJING model.
Although after averaging over many events only the magnetic field perpendicular to the reaction
plane is sizable, we find very strong electric and magnetic fields both parallel and
perpendicular to the reaction plane on the event-by-event basis. We study the time evolution
and the spatial distribution of these fields. Especially, the electromagnetic response of the quark-gluon
plasma
can give non-trivial evolution of the electromagnetic fields. The implications of the strong electromagnetic fields
on the hadronic observables are also discussed.
\end{abstract}

\maketitle

\section {Introduction}\label{intro}
Relativistic heavy-ion collisions provide us the methods to create and explore strongly interacting
matter at high energy densities where the deconfined quark-gluon plasma (QGP) is
expected to form. The properties of matter governed by quantum chromodynamics (QCD)
have been studied at the Relativistic Heavy Ion Collider (RHIC) at Brookhaven National Laboratory
(BNL) and at the Large Hadron Collider (LHC) at CERN. Measurements
performed at RHIC in Au + Au collisions at center-of-mass energy $\sqrt{s}=200$ GeV per nucleon
pair and at LHC in Pb + Pb collisions at center-of-mass energy $\sqrt{s}=2.76$ TeV per nucleon
pair have revealed several unusual properties of this hot, dense, matter (\eg, its very low shear
viscosity~\cite{arXiv:0804.4015,arXiv:1011.2783}, and its high opacity for energetic
jets~\cite{Wang:2003mm,Vitev:2002pf,Eskola:2004cr,Turbide:2005fk}).

Due to the fast, oppositely directed, motion of two colliding ions, off-central heavy-ion collisions can create
strong transient magnetic fields~\cite{Rafelski:1975rf}. As estimated by Kharzeev, McLerran, and
Warringa~\cite{arXiv:0711.0950},
the magnetic fields generated in off-central Au + Au collision at RHIC can reach
$eB\sim m_\p^2\sim 10^{18}$ G, which is $10^{13}$ times larger than the strongest man-made steady magnetic field
in the laboratory. The magnetic field generated at LHC energy can be $10$ times larger
than that at RHIC~\cite{arXiv:0907.1396}. Thus, heavy-ion collisions provide a unique
terrestrial environment to study QCD in strong magnetic fields. It has been shown that a strong magnetic field
can convert topological charge fluctuations in the QCD vacuum into global electric charge separation
with respect to the reaction plane~\cite{arXiv:0711.0950,arXiv:0808.3382,arXiv:0911.3715}.
This so-called {\it chiral magnetic effect} may serve as a sign
of the local P and CP violation of QCD. Experimentally, the STAR~\cite{arXiv:0909.1717,arXiv:0909.1739},
PHENIX~\cite{Ajitanand:2010rc}, and ALICE~\cite{collaboration:2011sma} Collaborations have reported the measurements
of the two-particle correlations of charged particles with respect to the reaction
plane, which are
qualitatively consistent with the chiral magnetic effect, although there are still some
debates~\cite{arXiv:1011.6053,arXiv:0912.5050,arXiv:1008.4919,arXiv:1101.1701,arXiv:0911.1482}.

Besides the chiral magnetic effect, there can be other effects caused by the strong magnetic fields including
the catalysis of chiral symmetry breaking~\cite{hep-ph/9405262}, the possible splitting of
chiral and deconfinement phase transitions~\cite{arXiv:1004.2712}, the spontaneous electromagnetic
superconductivity of QCD vacuum~\cite{arXiv:1008.1055,arXiv:1101.0117}, the possible enhancement of elliptic
flow of charged particles~\cite{arXiv:1102.3819,arXiv:1108.4394}, the energy loss due to the
synchrotron radiation of quarks~\cite{arXiv:1006.3051}, the emergence of anisotropic viscosities
~\cite{arXiv:1108.4394,arXiv:0910.3633,arXiv:1108.0602}, the induction of the electric quadrupole moment of the
QGP~\cite{arXiv:1103.1307}, \etc.

\begin{figure}[!htb]
\begin{center}
\includegraphics[width=7cm]{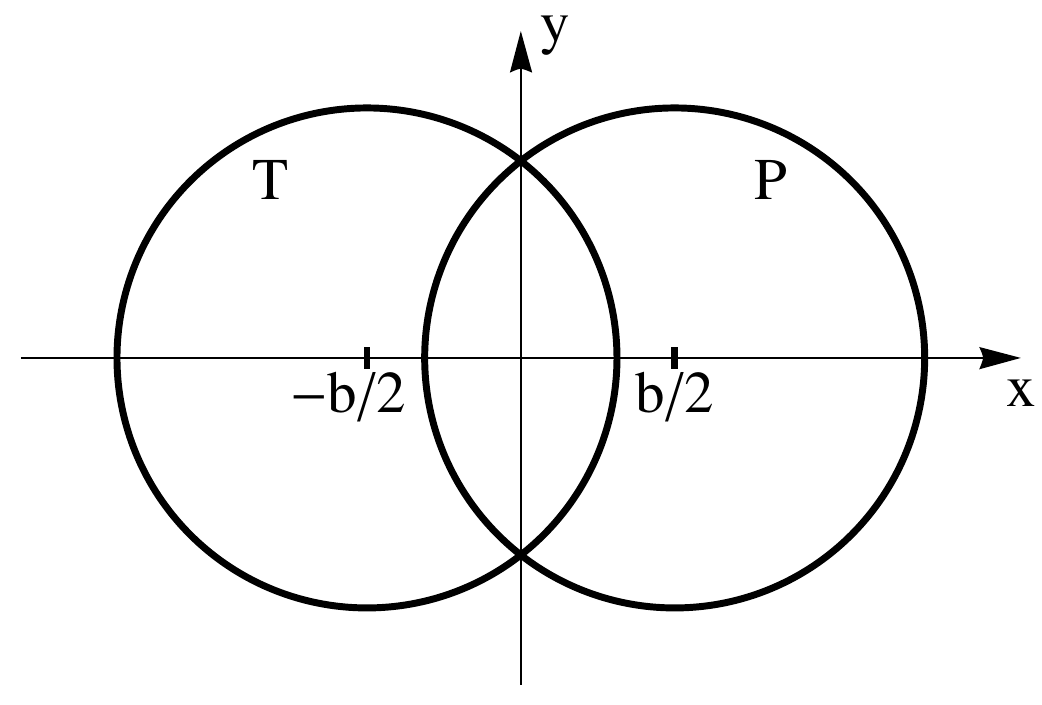}
\caption{The geometrical illustration of the off-central collisions with impact parameter $b$.
Here ``T" for target and ``P" for projectile.}
\label{collision}
\end{center}
\end{figure}

The key quantity of all these effects are the strength of the magnetic fields.
Most of the previous works estimated the magnetic-field
strength based on the averaging over many events
~\cite{arXiv:0711.0950,arXiv:0907.1396,arXiv:1003.2436,arXiv:1103.4239,arXiv:1107.3192},
thus, due to the mirror
symmetry of the collision geometry, only the $y$-component of the magnetic field remains sizable,
$e\lan B_y\ran\sim m_\p^2$, while other components are $\lan B_x\ran=\lan B_z\ran=0$.
Hereafter, we use the angle bracket to denote event average. We choose the $z$ axis along the beam direction of
the projectile, $x$ axis along the impact parameter ${\bf b}$
from the target to the projectile, and $y$ axis perpendicular to
the reaction plane, as illustrated in \fig{collision}.

However, in many cases, the final hadronic signals are measured on the event-by-event basis.
Thus, it is important
to study how the magnetic fields are generated on the event-by-event basis.
Such a study was recently initiated by Bzdak and Skokov~\cite{arXiv:1111.1949}. They showed that
the $x$-component of the magnetic field can be as strong as the $y$ component on the event-by-event basis,
due to the fluctuation of the proton positions in the
colliding nuclei.
Besides, they also found that the event-by-event generated electric field can be comparable to the
magnetic field.

The aim of our work is to give a detailed study of the space-time structure of the event-by-event generated
electromagnetic fields in the heavy-ion collisions.
We perform our calculation by using the heavy ion jet interaction generator (HIJING) model~\cite{Wang:1991hta,Deng:2010mv,Deng:2010xg}.
HIJING is a Monte-Carlo event generator for hadron productions
in high energy p + p, p + A, and A + A collisions. It is essentially a two-component model, which describes the
production of hard parton jets and the soft interaction between nucleon remnants. In HIJING, the hard jets production
is controlled by perturbative QCD, and the interaction of nucleon remnants via soft gluon exchanges is described
by the string model \cite{Sjostrand:1987su}.

This paper is organized as follows. We give a general setup of our calculation in Sec.~\ref{gener}.
In Sec.~\ref{resul}, we
present our numerical results. We discuss the influence of the non-trivial electromagnetic response of the QGP on the time evolution
of the electromagnetic fields in Sec.~\ref{respo}. We conclude with discussions and summary in Sec.~\ref{discu}. We use natural unit $\hbar=c=1$.

\section {General Setup}\label{gener}
We use the Li\'enard-Wiechert potentials to calculate the electric and magnetic fields at a position $\br$ and time $t$,
\begin{eqnarray}
\label{LWE}
e\bE(t,\br)&=&\frac{e^2}{4\p}\sum_n Z_n\frac{{\bf R}_n-R_n\bv_n}{(R_n-{\bf R}_n\cdot\bv_n)^3}(1-v_n^2),\\
\label{LWB}
e\bB(t,\br)&=&\frac{e^2}{4\p}\sum_n Z_n\frac{\bv_n\times{\bf R}_n}{(R_n-{\bf R}_n\cdot\bv_n)^3}(1-v_n^2),
\end{eqnarray}
where $Z_n$ is the charge number of the $n$th particle, ${\bf R}_n=\br-\br_n$ is the relative
position of the field point $\br$ to the source
point $\br_n$, and $\br_n$ is the location of the $n$th particle with velocity $\bv_n$ at the retarded time
$t_n=t-|\br-\br_n|$. The summations run over all
charged particles in the system. Although there are singularities at $R_n=0$ in \eqs{LWE}{LWB}, in practical
calculation of $\bE$ and $\bB$ at given $(t,\br)$, the events causing such singularities rarely appear.
So, we omit such events in our numerical code.
In non-relativistic limit, $v_n\ll1$, \eq{LWE} reduces to
the Coulomb's law and \eq{LWB} reduces to the Biot-Savart law for a set of moving charges,
\begin{eqnarray}
\label{LWE2}
e\bE(t,\br)&=&\frac{e^2}{4\p}\sum_n Z_n\frac{{\bf R}_n}{R^3_n},\\
\label{LWB2}
e\bB(t,\br)&=&\frac{e^2}{4\p}\sum_n Z_n\frac{\bv_n\times{\bf R}_n}{R^3_n}.
\end{eqnarray}

To calculate the electromagnetic fields at moment $t$,
we need to know the full phase space information of all charged particles before $t$. In the HIJING model,
the position of each nucleon before collision is sampled according to the Woods-Saxon distribution.
The energy for each nucleon is set to be $\sqrt{s}/2$ in the
center-of-mass frame. Assuming that the nucleons have no transverse momenta before collision, the value
of the velocity of each nucleon is given by $v_z^2=1-(2m_N/\sqrt{s})^2$, where $m_N$ is the mass of the nucleon.
At RHIC and LHC, $v_z$ is very large, so the nuclei are Lorentz contracted to pancake shapes.

We set the initial time $t=0$ as the moment when the two nuclei completely overlap. The collision time
for each nucleon is given according to its initial longitudinal position $z_N^L=z_N\cdot2m_N/\sqrt{s}$
and velocity $v_z$, where $z_N$ is the initial longitudinal position of nucleon in the rest frame of the nucleus.
The probability of two nucleon colliding at a given impact parameter $b$ is determined by the Glauber
model. In this paper, we call nucleons without any interaction {\it spectators} and those that suffer at
least once elastic or inelastic collision {\it participants} before their first collision. The participants will exchange
their momenta and energies, and become {\it remnants} after collision. Differently from the spectators and
the participants, the remnants can have finite transverse momenta.
In our calculation, we find that although the spectators and participants are the main sources of
fields at $t\leq0$, remnants can give important contributions at $t>0$.
In the HIJING model,
we neglect the back reaction of the electromagnetic field
on the motions of the the charged particles. This is a good approximation before the collision happens because
the electromagnetic field is weak at that time. We will discuss
the feed back effect of the electromagnetic fields on QGP in Sec. \ref{respo} by using a
magnetohydrodynamic treatment.

After collision, many partons may be produced and the hot, dense, QGP may form. As the QGP is nearly neutral,
we neglect the contributions from the produced partons to the generation of the electromagnetic field.
However, if the electric
conductivity of the QGP is large, the QGP can have significant response to the change of the
external electromagnetic field. This can become substantial for the time evolution of the fields in the QGP. We will
discuss this point in detail in Sec. \ref{respo}.

\section {Numerical Results}\label{resul}
\subsection {Impact parameter dependence}\label{subsec_bd}
\begin{figure}[!htb]
\begin{center}
\includegraphics[width=7.0cm]{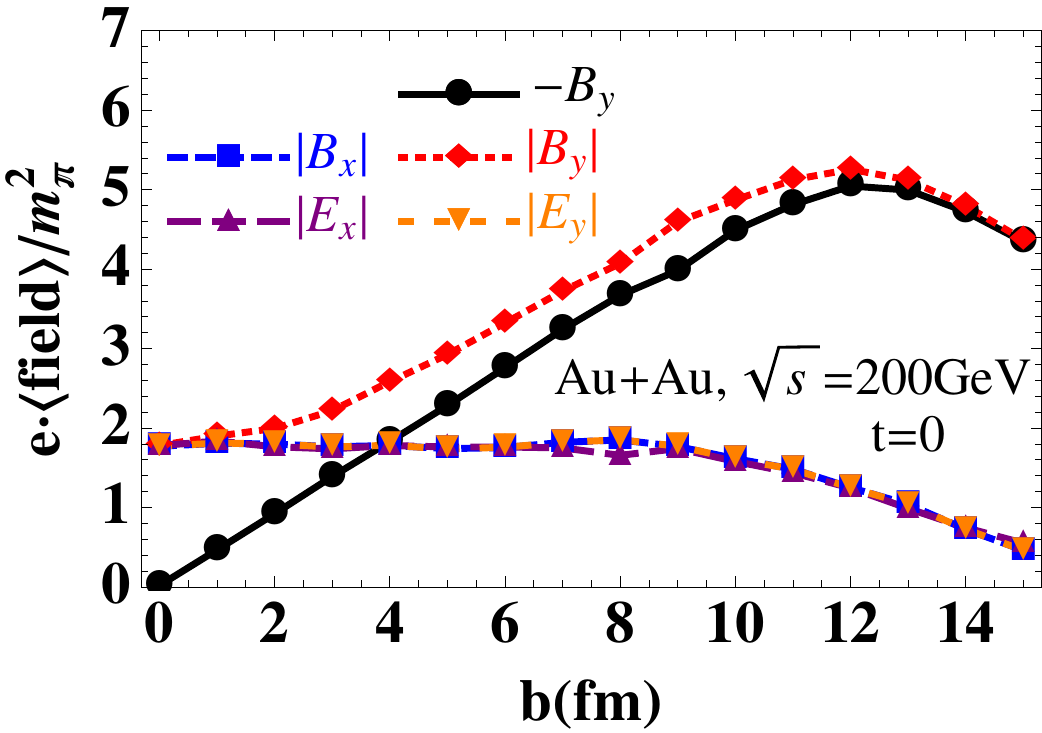}
\includegraphics[width=7.0cm]{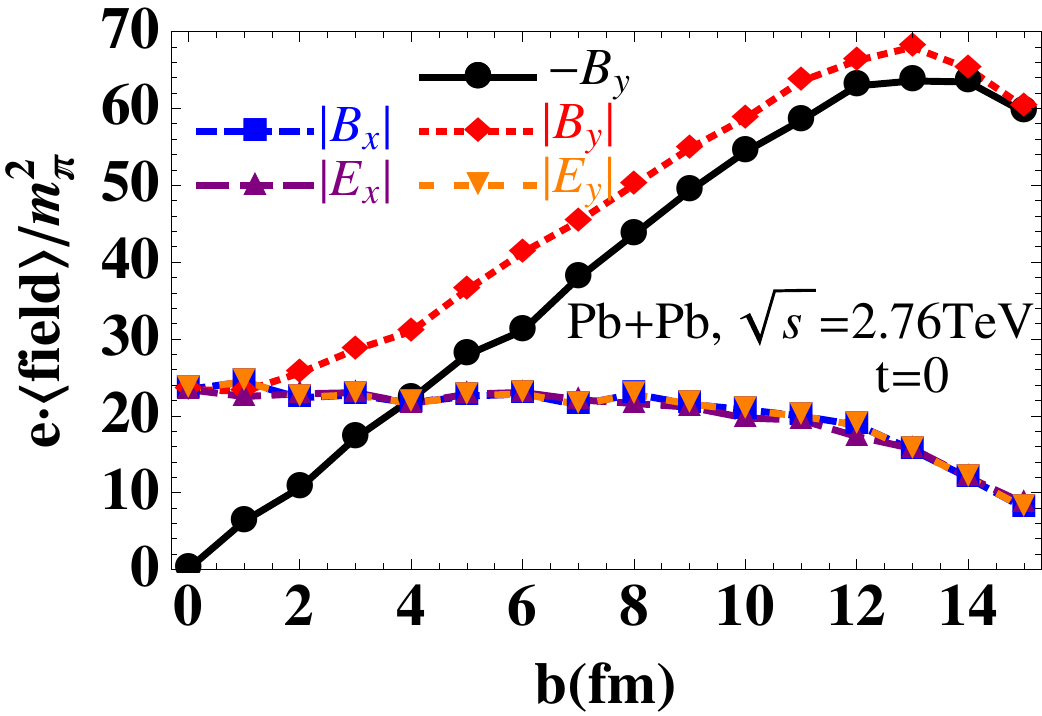}
\caption{(Color online) The electromagnetic fields at $t=0$ and $\br={\bf 0}$
as functions of the impact parameter $b$.}
\label{bdepe}
\end{center}
\end{figure}

We first show the impact parameter dependence of the electromagnetic fields at $\br={\bf 0}$ and $t=0$.
The left panel of \fig{bdepe} is the results for Au + Au collision at RHIC energy $\sqrt{s}=200$
GeV; the right panel of \fig{bdepe} is for Pb + Pb collision at LHC energy $\sqrt{s}=2.76$ TeV.
As seen from \eq{LWB}, $\lan B_x(t,{\bf 0})\ran=0$, while $\lan B_y(t,{\bf 0})\ran<0$
when $b>0$. Also, from \eqs{LWE}{LWB}, we find that there are always
$|E_y(0,{\bf 0})|\approx|B_x(0,{\bf 0})|$ and $|B_y(0,{\bf 0})|\geq |E_x(0,{\bf 0})|$
when $v_z$ is large [See \eqs{et0}{bt0}].
These facts are reflected in \fig{bdepe}. Although the $x$-component of the magnetic
field as well as the $x$- and $y$-components of the electric field vanish after averaging
over many events, their magnitudes
in each event can be huge due to the fluctuations of the proton positions in the nuclei.
Thus, following Bzdak and Skokov~\cite{arXiv:1111.1949}, we plot the averaged absolute values
$\lan|E_{x,y}|\ran$ and $\lan|B_{x,y}|\ran$ at $\br={\bf 0}$ and $t=0$.
Similar with the findings in Ref.~\cite{arXiv:1111.1949}, we find that $\lan|B_x|\ran$,
$\lan|E_x|\ran$, and $\lan|E_y|\ran$ are comparable to $\lan|B_y|\ran$, and the following
equalities hold approximately, $\lan|E_x|\ran\approx\lan|E_y|\ran\approx\lan|B_x|\ran$.
But our results at RHIC energy are about three times smaller than that obtained in  Ref.~\cite{arXiv:1111.1949}.
We checked that this is because the thickness of the nuclei in our calculation is finite while
the authors of Ref.~\cite{arXiv:1111.1949} assumed that the nuclei are infinitely thin.
We can also observe that, at small $b$ region, contrary to $\lan B_y\ran$ which is proportional to $b$,
the fields caused by fluctuation are not sensitive to $b$.

\subsection {Collision energy dependence}\label{subsec_ced}
\begin{figure*}[!htb]
\begin{center}
\includegraphics[width=6.5cm]{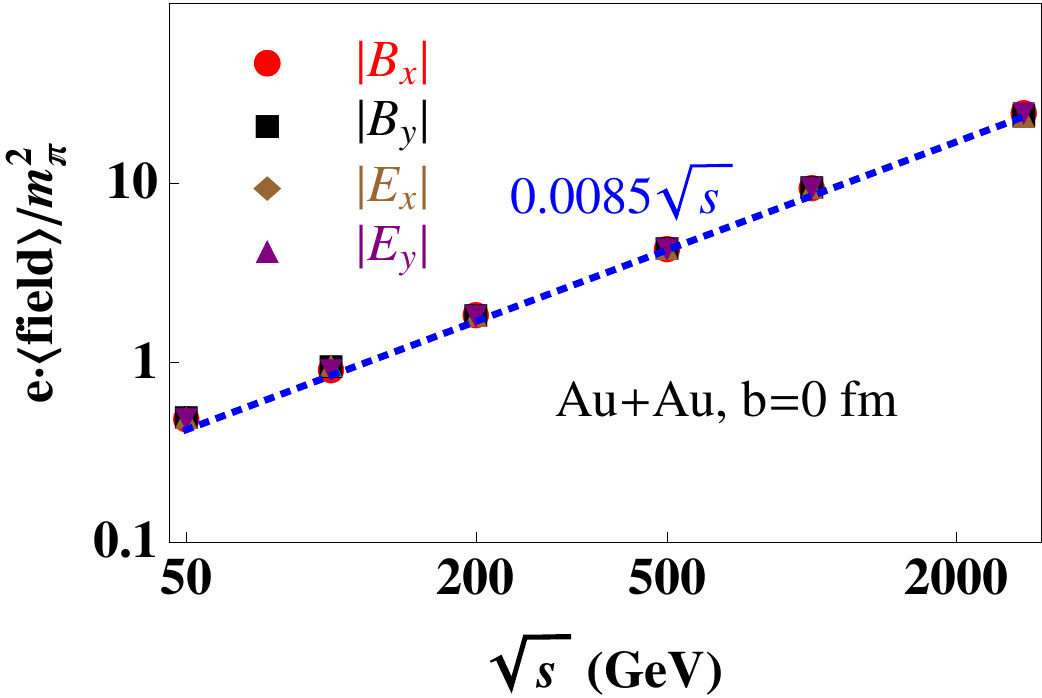}
\includegraphics[width=6.5cm]{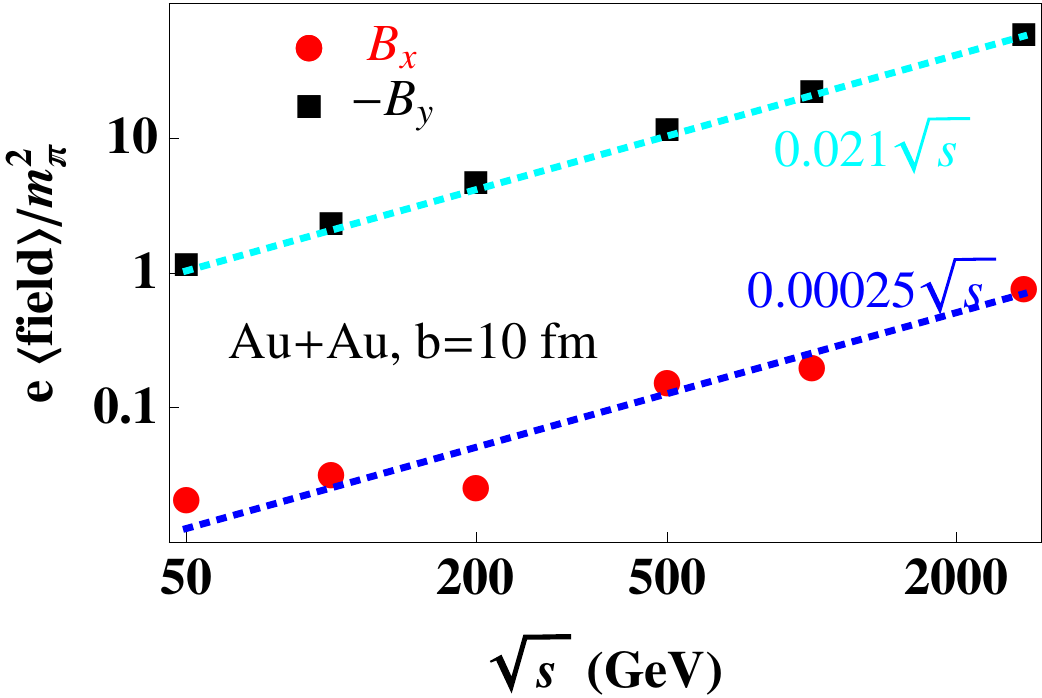}
\includegraphics[width=6.5cm]{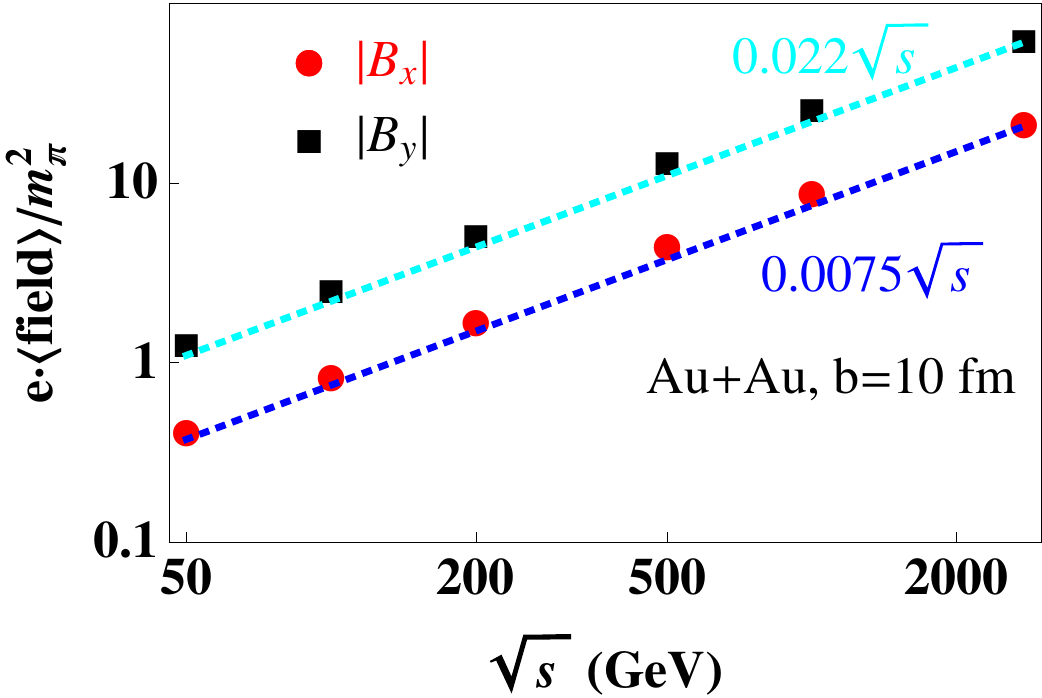}
\includegraphics[width=6.5cm]{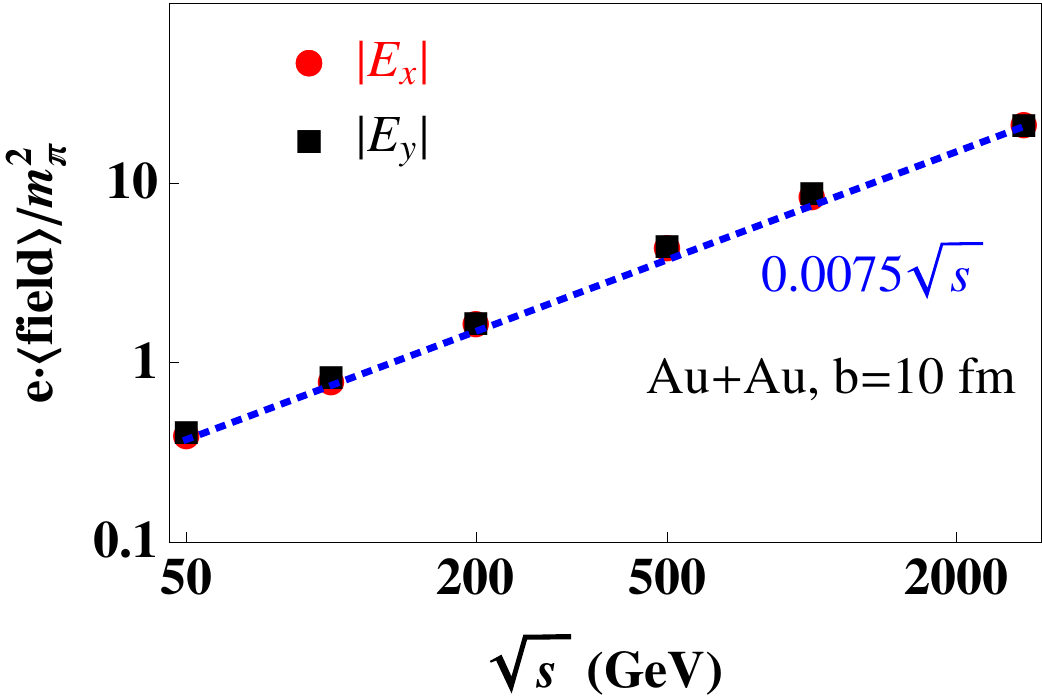}
\caption{(Color online) The collision energy dependence of the electromagnetic fields at $\br={\bf 0}$
and $t=0$.}
\label{cedepe}
\end{center}
\end{figure*}
We see from \fig{bdepe} that the magnitudes of all the fields at LHC energy is around
$14$ times bigger than that at RHIC energy. To study the collision energy dependence
more carefully, we calculate the fields at $t=0$ and $\br={\bf 0}$ for different $\sqrt{s}$.
To high precision, the linear dependence of the fields on the collision energy
is obtained, as shown in \fig{cedepe}. Thus, the following scaling law holds for
event-by-event generated electromagnetic fields as well as for event-averaged magnetic fields,
\begin{eqnarray}
\label{linear}
e\cdot{\rm Field}\propto\sqrt{s}f(b/R_A),
\end{eqnarray}
where $R_A$ is the radius of the nucleus and $f(b/R_A)$ is a universal function which
has the shapes as shown in \fig{bdepe} for $\lan |B_{x,y}|\ran$,
$\lan |E_{x,y}|\ran$, and $\lan B_y\ran$.

Actually, a more general form of \eq{linear} can be derived from \eqs{LWE}{LWB}.
As the fields at $t=0$ are mainly caused by spectators and participants whose velocity
$v_n=v_z=\sqrt{1-(2m_N/\sqrt{s})^2}\approx 1$,
the electric and magnetic fields at $t=0$ in the transverse plane can be expressed as
\begin{eqnarray}
\label{et0}
e\bE_\perp(0,{\bf r})&\approx&\frac{e^2}{4\p}\frac{\sqrt{s}}{2m_N}\sum_n \frac{{\bf R}_{n\perp}}
{R_{n\perp}^3},\\
\label{bt0}
e\bB_\perp(0,{\bf r})&\approx&\frac{e^2}{4\p}\frac{\sqrt{s}}{2m_N}\sum_n \frac{{\bf e}_{nz}\times{\bf R}_{n\perp}}
{R_{n\perp}^3},
\end{eqnarray}
where ${\bf e}_{nz}$ is the unit vector in $\pm z$ direction depending on whether the $n$th proton
is in the target or in the projectile, ${\bf R}_{n\perp}$ is the transverse position of
the $n$th proton which is independent of $\sqrt{s}$, and $R_{n\perp}=|{\bf R}_{n\perp}|$.

\subsection {Spatial distribution}\label{subsec_sd}
\begin{figure*}[!htb]
\begin{center}
\includegraphics[width=8.0cm]{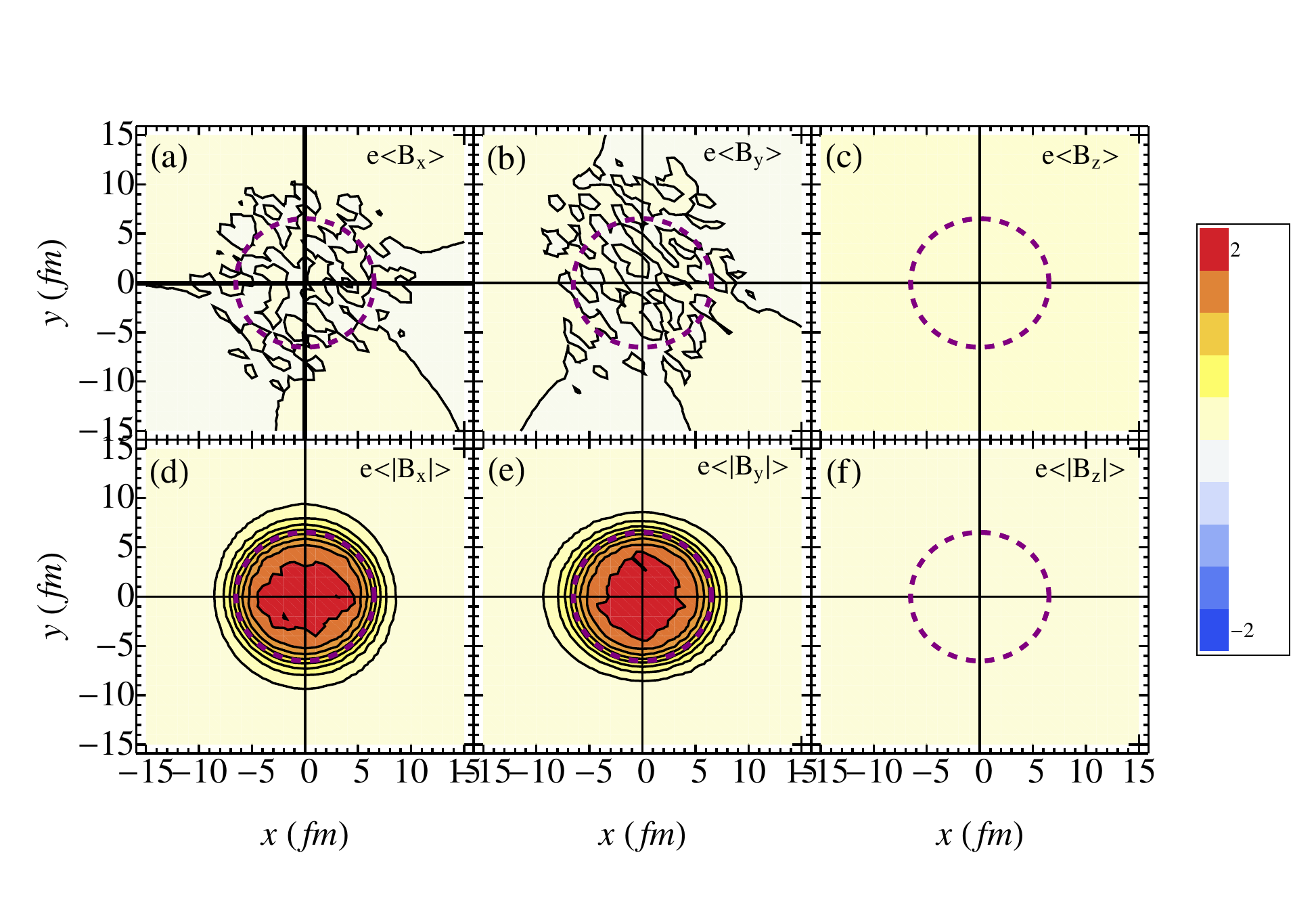}
\includegraphics[width=8.0cm]{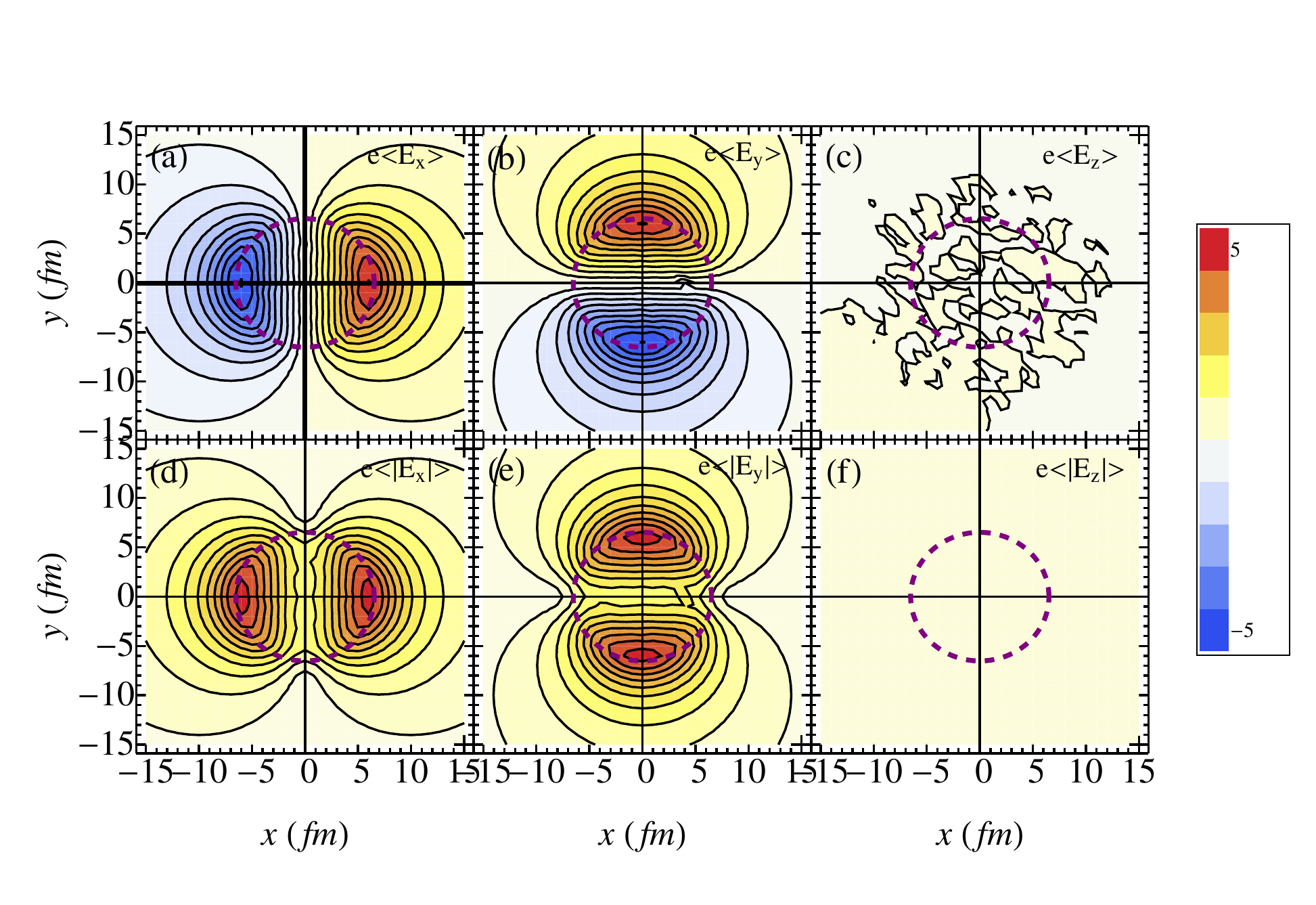}
\includegraphics[width=8.0cm]{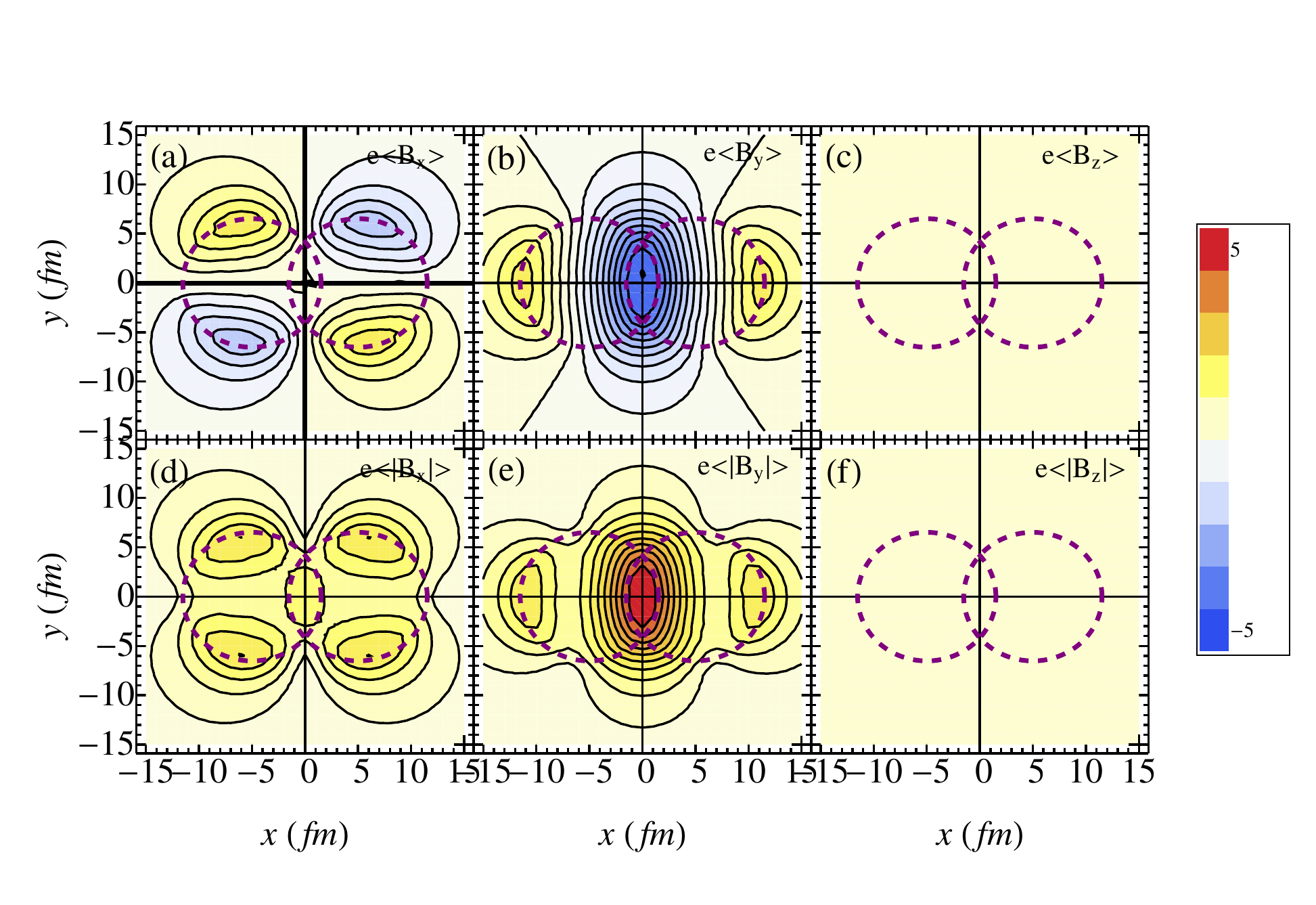}
\includegraphics[width=8.0cm]{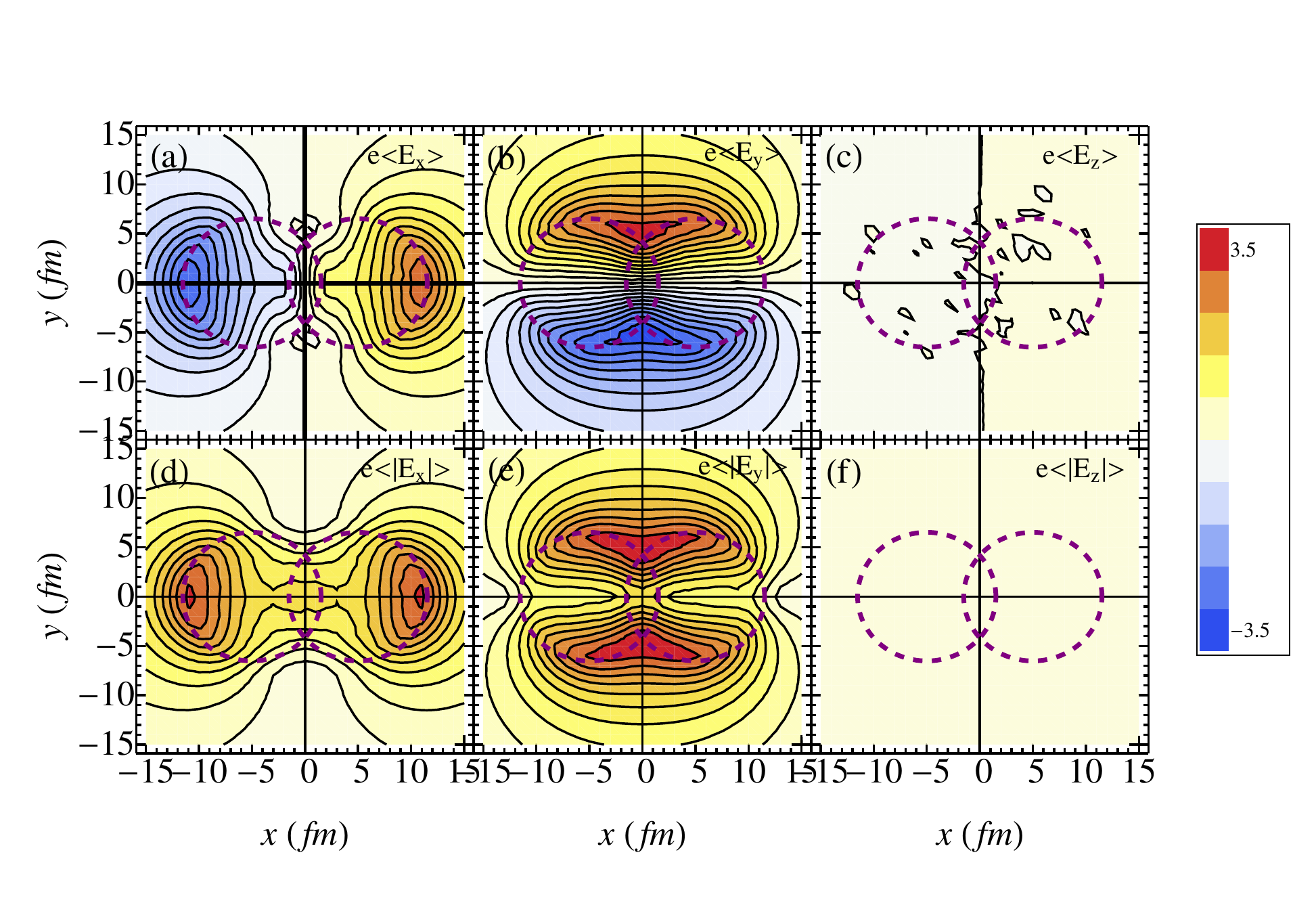}
\caption{(Color online) The spatial distributions of the electromagnetic fields in the transverse plane at $t=0$ for
$b=0$ (upper panels) and $b=10$ fm (lower panels) at RHIC energy.
The unit is $m_\p^2$. The dashed circles indicate the two colliding nuclei.}
\label{sdist}
\end{center}
\end{figure*}
The spatial distributions of the magnetic and electric fields are evidently inhomogeneous.
We show in \fig{sdist} the contour plots of $\lan B_{x,y,z}\ran$, $\lan E_{x,y,z}\ran$,
$\lan |B_{x,y,z}|\ran$, and $\lan |E_{x,y,z}|\ran$ at $t=0$ in the transverse plane at RHIC energy.
The upper two panels are for $b=0$ and the lower two panels are for $b=10$ fm. The spatial
distribution of the transverse fields for LHC energy is merely the same as \fig{sdist} but the fields have
$2760/200\approx 14$ times larger magnitudes everywhere according to \eqs{et0}{bt0}.
The spatial distribution of the fields in the reaction plane was studied in Ref.~\cite{arXiv:1103.4239}.

First, as we expected, the longitudinal fields $\lan B_z\ran$, $\lan E_z\ran$,
$\lan|B_z|\ran$, and $\lan|E_z|\ran$ are much smaller than the transverse fields.
Second, the event-averaged fields $\lan B_{x,y}\ran$ and $\lan E_{x,y}\ran$ distribute
similarly with the fields generated by two uniformly charged, oppositely moving, discs.
Third, the spatial distribution of the magnetic fields is very different
from that of the electric fields on the event-by-event basis.
For central collisions, both $\lan|B_{x}|\ran$ and $\lan|B_{y}|\ran$ distribute circularly and
concentrate at $\br={\bf 0}$, while $\lan |E_{x}|\ran$ and $\lan |E_{y}|\ran$ peak around $x=\pm R_A$ and $y=\pm R_A$
with $R_A$ the radius of the nucleus. We notice that for off-central collision, the $y$-component of
the electric field varies steeply along $y$-direction, reflecting the fact that at $t=0$ a large amount
of net charge stays temporally in the ``almond''-shaped overlapping region.

\subsection {Probability distribution over events}\label{subsec_pd}
\begin{figure*}[!htb]
\begin{center}
\includegraphics[width=7.5cm]{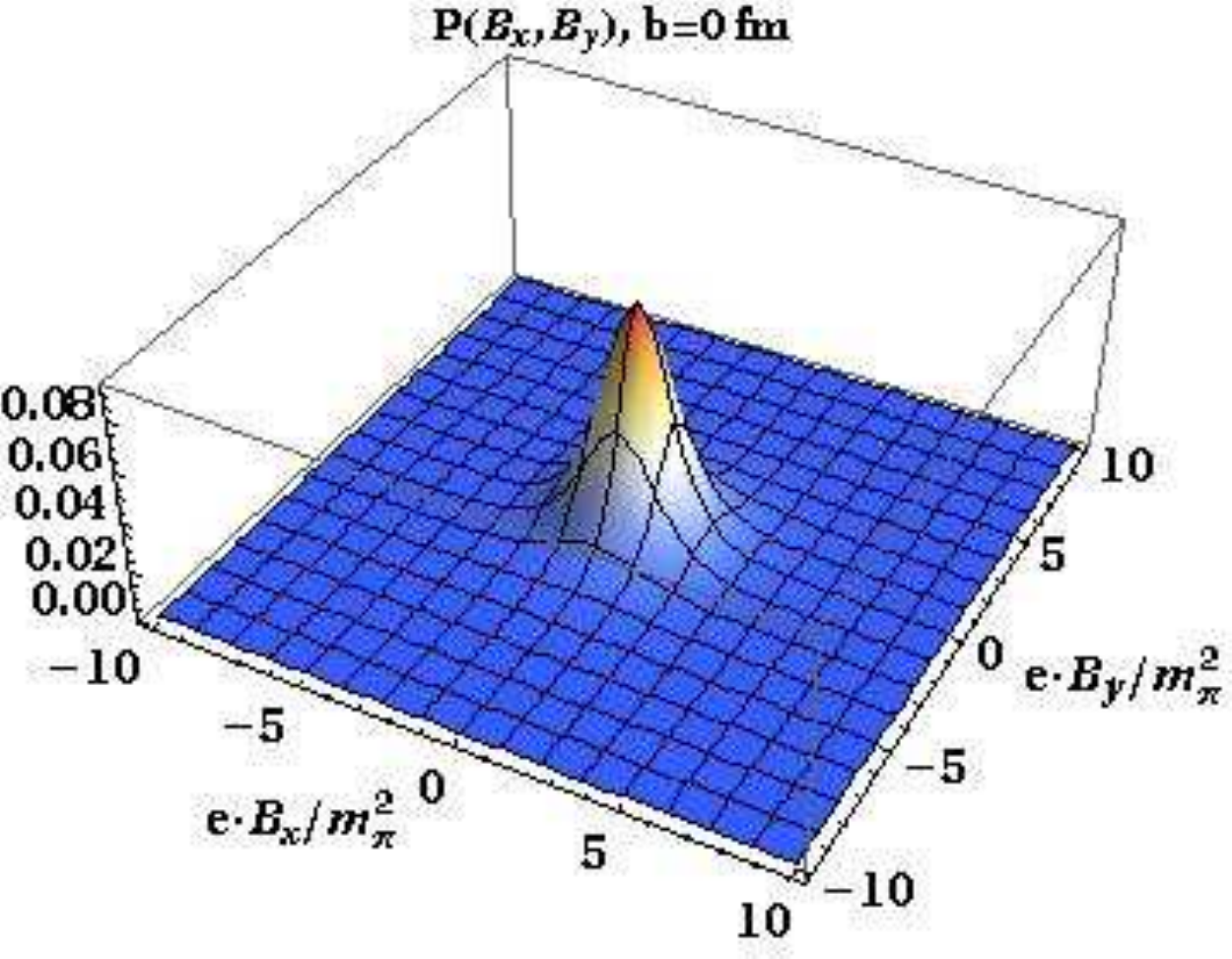}
\includegraphics[width=7.5cm]{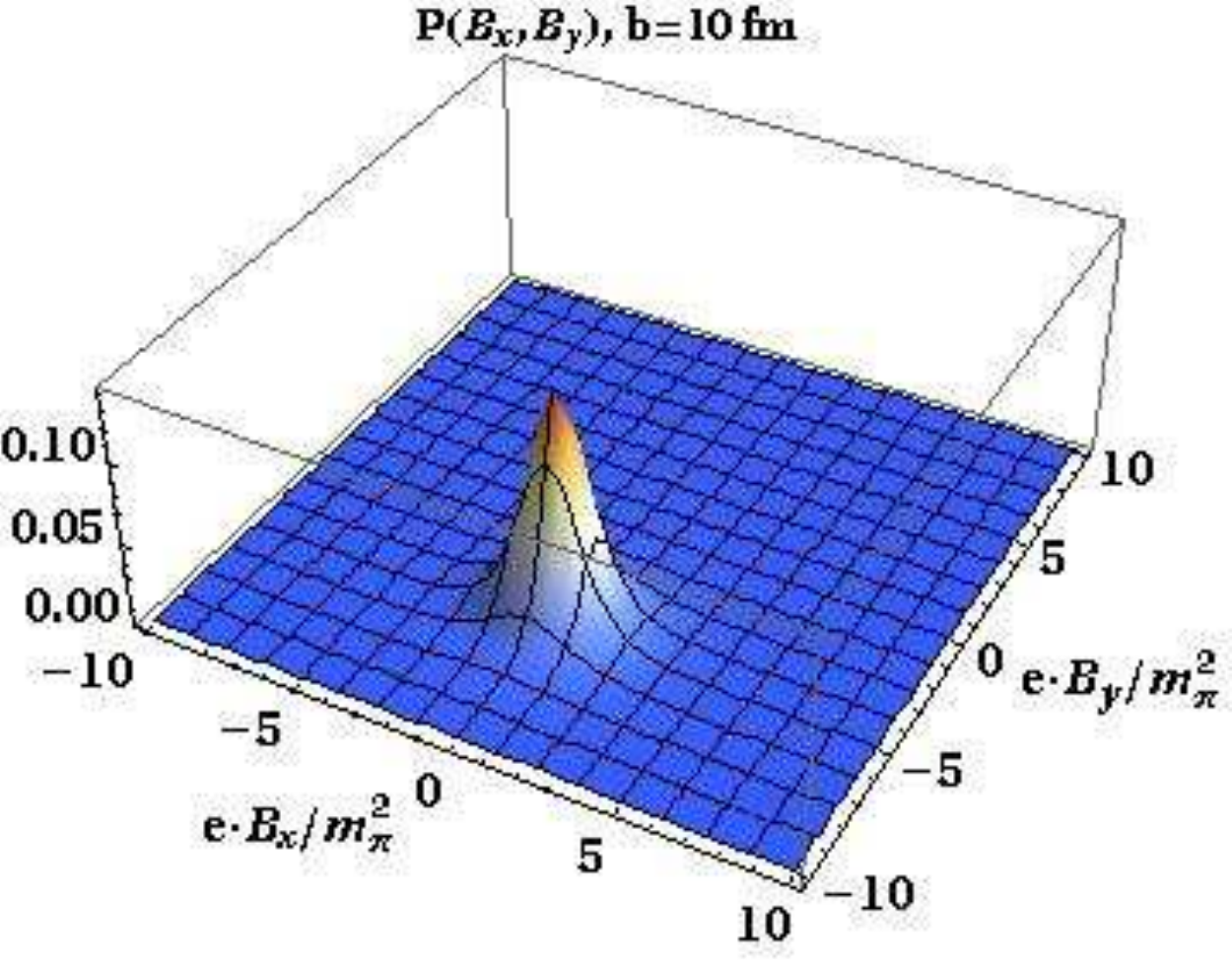}
\includegraphics[width=7.5cm]{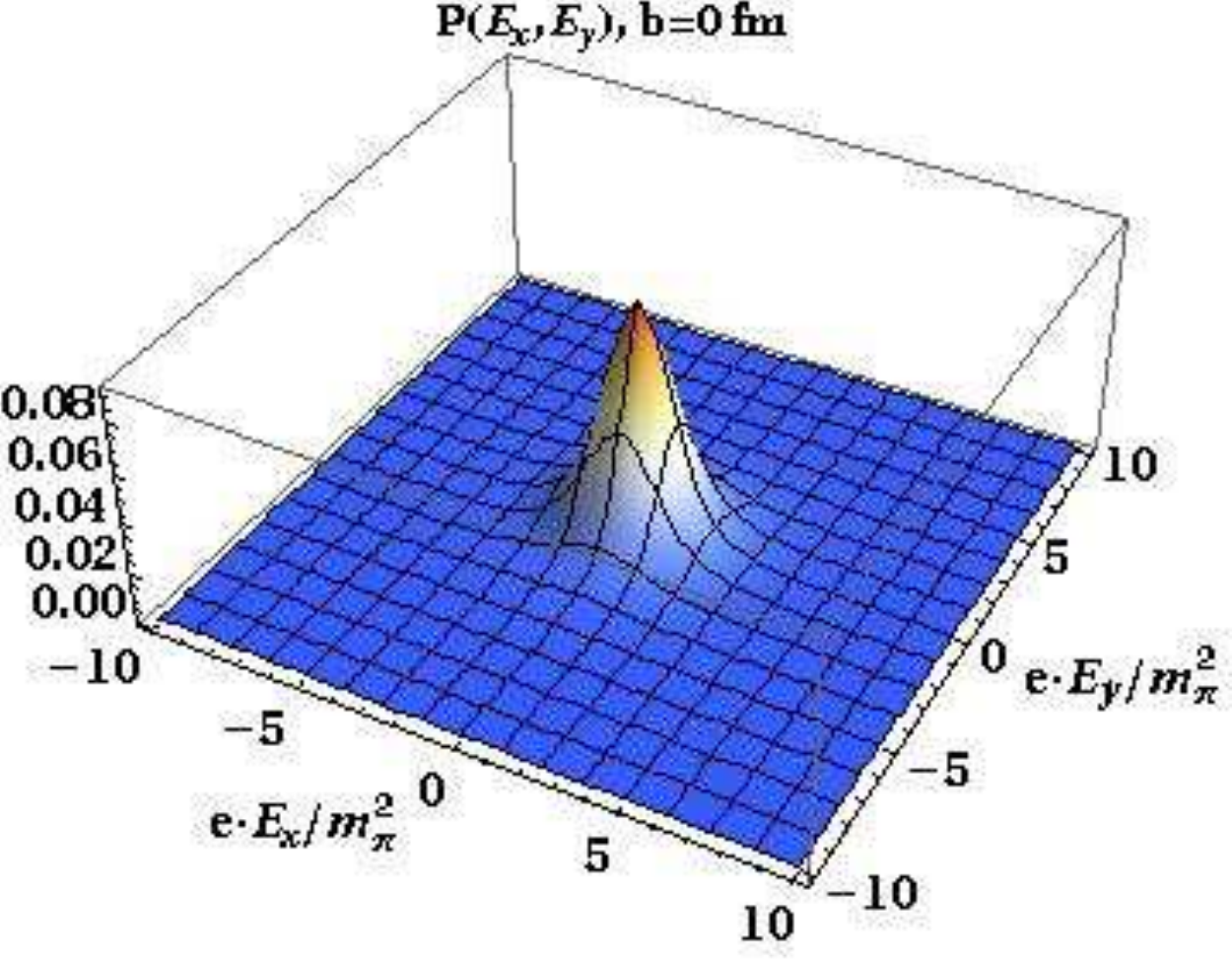}
\includegraphics[width=7.5cm]{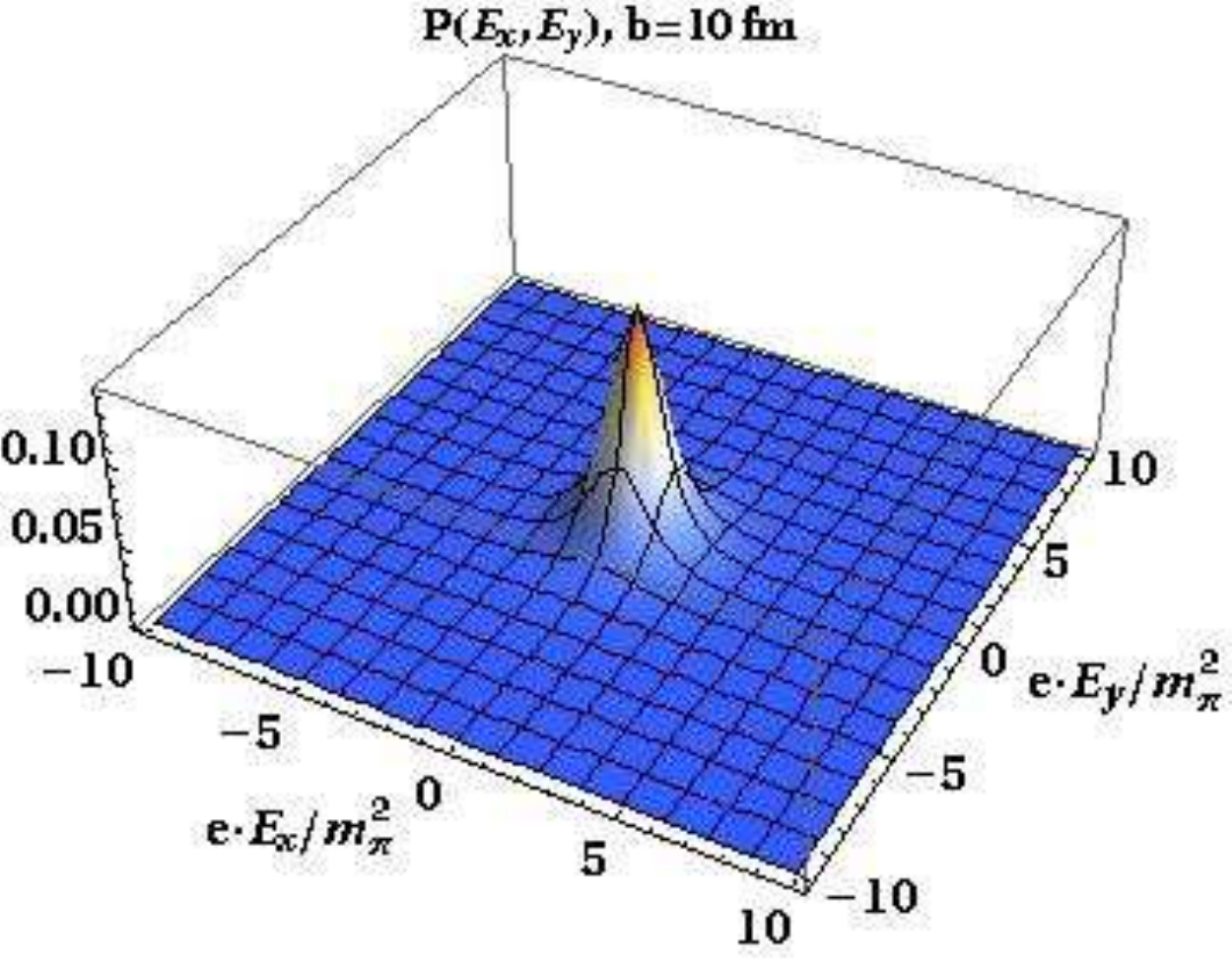}
\caption{(Color online) The probability densities $P(B_x, B_y)$ and $P(E_x,E_y)$
for different impact parameters for Au + Au collisions at $\sqrt{s}=200$ GeV.}
\label{p3d}
\end{center}
\end{figure*}
\begin{figure*}[!htb]
\begin{center}
\includegraphics[width=7.5cm]{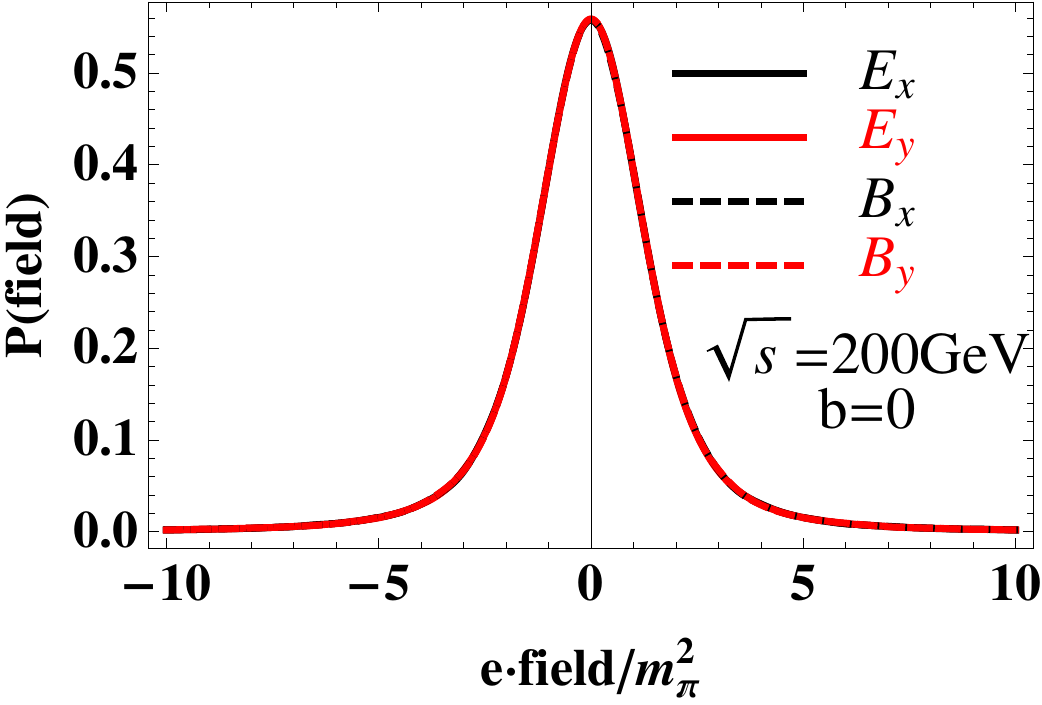}
\includegraphics[width=7.5cm]{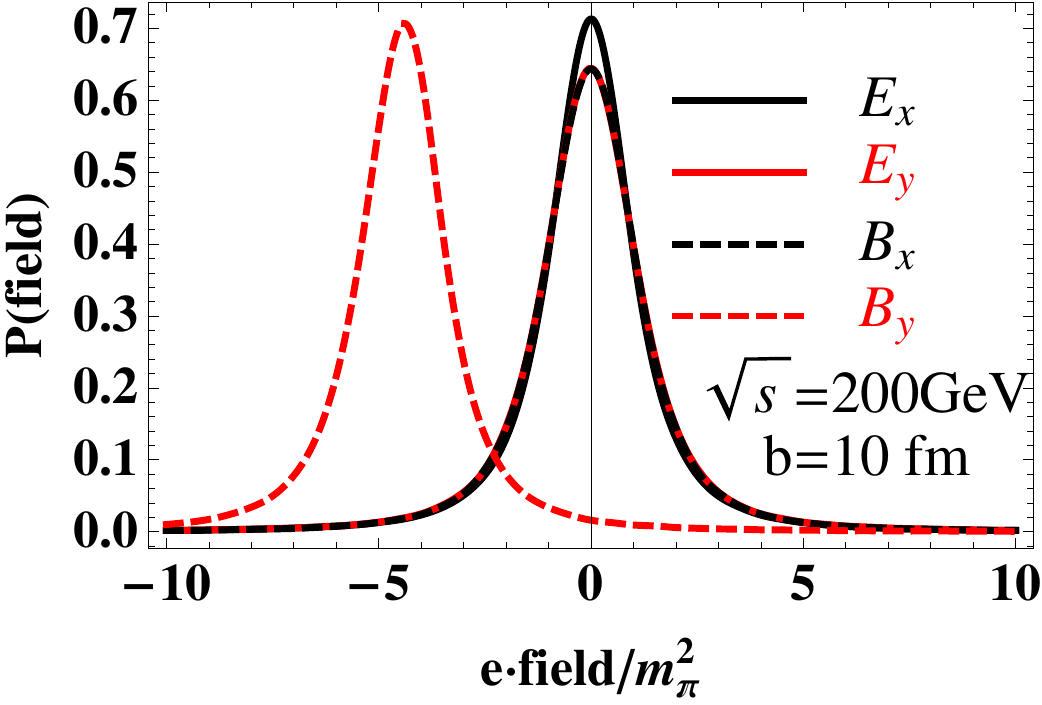}
\includegraphics[width=7.5cm]{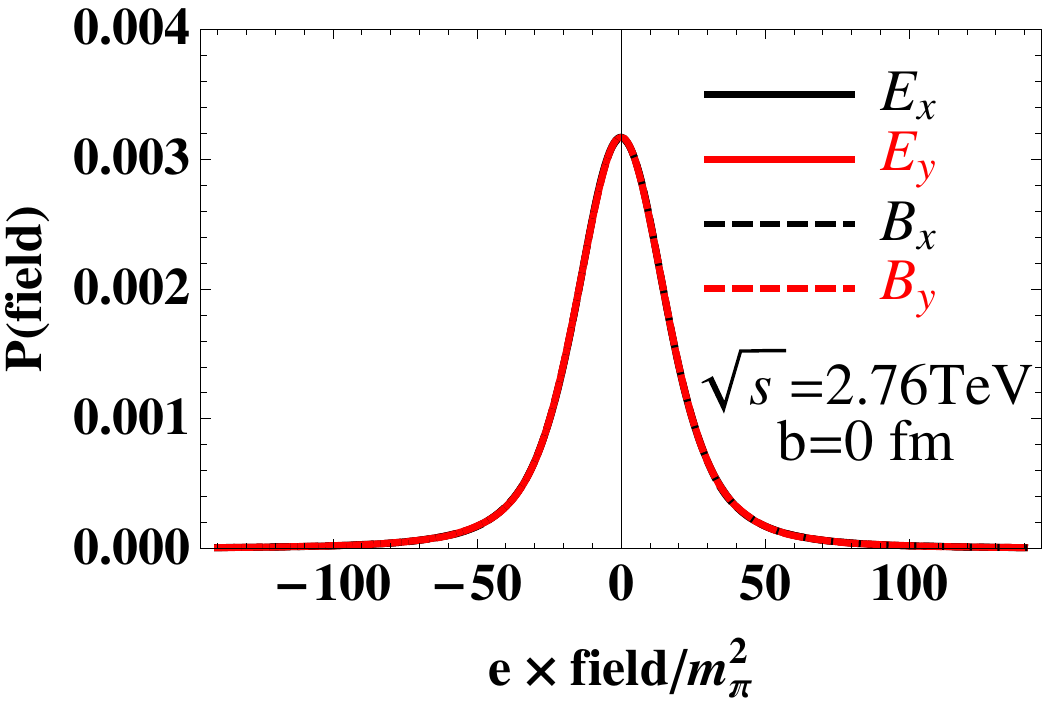}
\includegraphics[width=7.5cm]{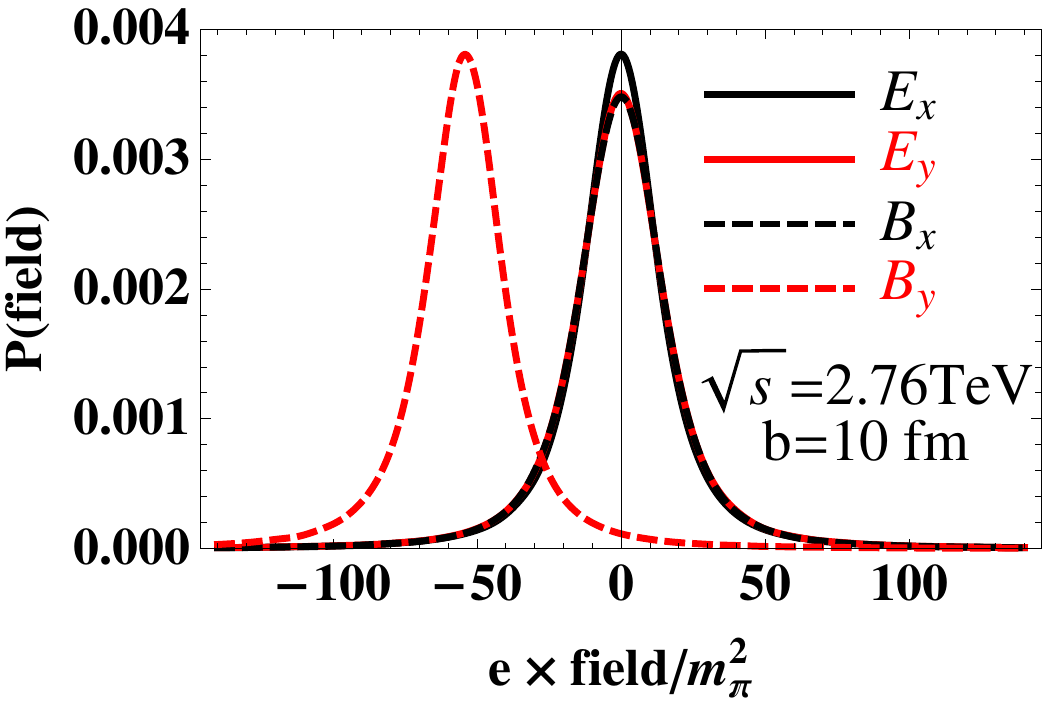}
\caption{(Color online) The probability densities $P(B_{x,y})$ and $P(E_{x,y})$ for central collision $b=0$
and off-central collision $b=10$ fm for Au + Au collision (upper panels) at $\sqrt{s}=200$ GeV and for Pb + Pb collision at $\sqrt{s}=2.76$ TeV (lower panels).}
\label{p2d}
\end{center}
\end{figure*}

Although we used the event-averaged absolute values $\lan |B_{x,y}|\ran$ and $\lan |E_{x,y}|\ran$ to
characterize the event-by-event fluctuations of the electromagnetic fields,
it would have more practical relevance to see the probability distribution of the magnetic field, defined as
\begin{eqnarray}
\label{pbb}
P(B_x,B_y)&\equiv&\frac{1}{N}\frac{d^2N}{dB_xdB_y},
\end{eqnarray}
where $N$ is the number of events. Similarly, we can define $P(E_x,E_y)$. After simulating $10^6$ events,
we obtain $P(B_x,B_y)$ and $P(E_x,E_y)$ at $t=0$ and $\br={\bf 0}$ for Au + Au collisions at $\sqrt{s}=200$ GeV,
as shown in \fig{p3d}.
As expected, the probability distribution of the magnetic (electric) field
peaks at $\bB={\bf 0}$ ($\bE={\bf 0}$) for central collision, while the probability distribution for magnetic field is shifted to finite
$B_y$ for off-central collision. This is more clearly shown in \fig{p2d}, where we depict
the one-dimensional probability density
$P(B_x)\equiv\int dB_y P(B_x,B_y)$ (other probability densities are analogously defined).

The probability distributions for Pb + Pb collisions at $\sqrt{s}=2.76$ TeV have analogous
shapes with \fig{p3d} but much more spread, as clearly shown in the lower panels of \fig{p2d}.
This is because the strength of the field generated at LHC can be obtained approximately from
that at RHIC by a $\sqrt{s_{\rm LHC}/s_{\rm RHIC}}$-scaling according to \eqs{et0}{bt0}. Thus, after normalization,
the probability distributions at LHC energy are related to that at RHIC energy by
\begin{eqnarray}
P_{\rm LHC}(B_x,B_y)\approx\frac{s_{\rm RHIC}}{s_{\rm LHC}}
P_{\rm RHIC}\lb\sqrt{\frac{s_{\rm RHIC}}{s_{\rm LHC}}}B_x,\sqrt{\frac{s_{\rm RHIC}}{s_{\rm LHC}}}B_y\rb,\\
P_{\rm LHC}(E_x,E_y)\approx\frac{s_{\rm RHIC}}{s_{\rm LHC}}
P_{\rm RHIC}\lb\sqrt{\frac{s_{\rm RHIC}}{s_{\rm LHC}}}E_x,\sqrt{\frac{s_{\rm RHIC}}{s_{\rm LHC}}}E_y\rb.
\end{eqnarray}

\subsection {Early-stage time evolution}\label{subsec_td}
\begin{figure*}[!htb]
\begin{center}
\includegraphics[width=6.5cm]{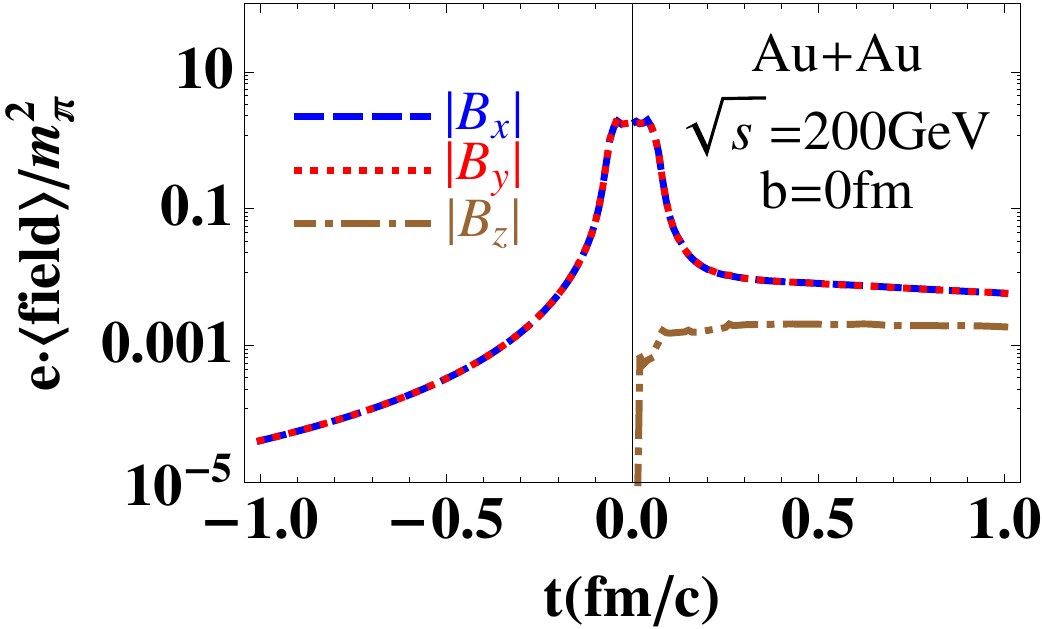}
\includegraphics[width=6.5cm]{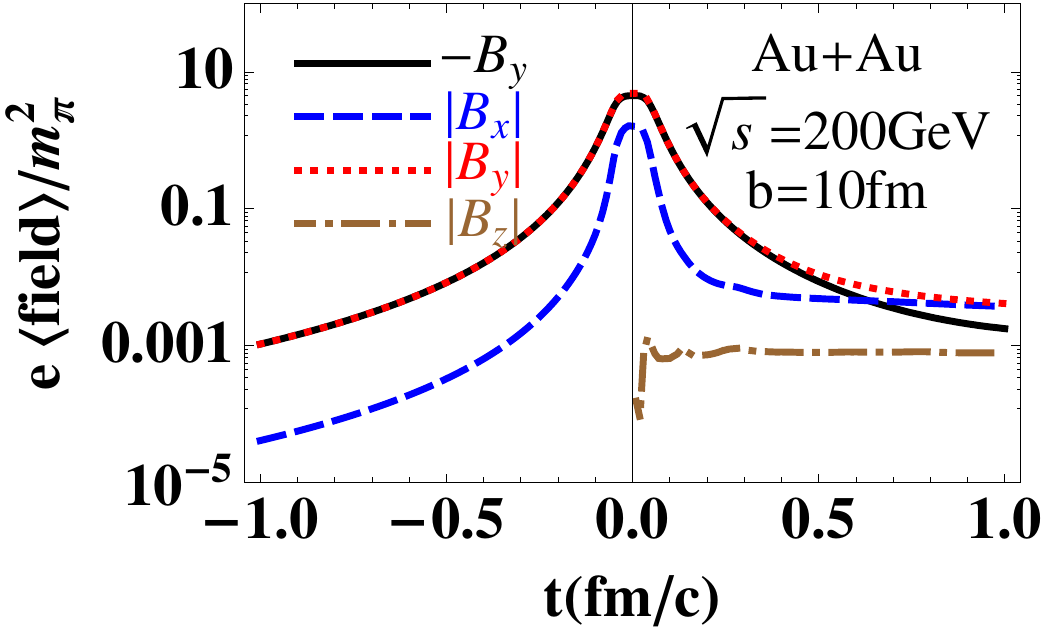}
\includegraphics[width=6.5cm]{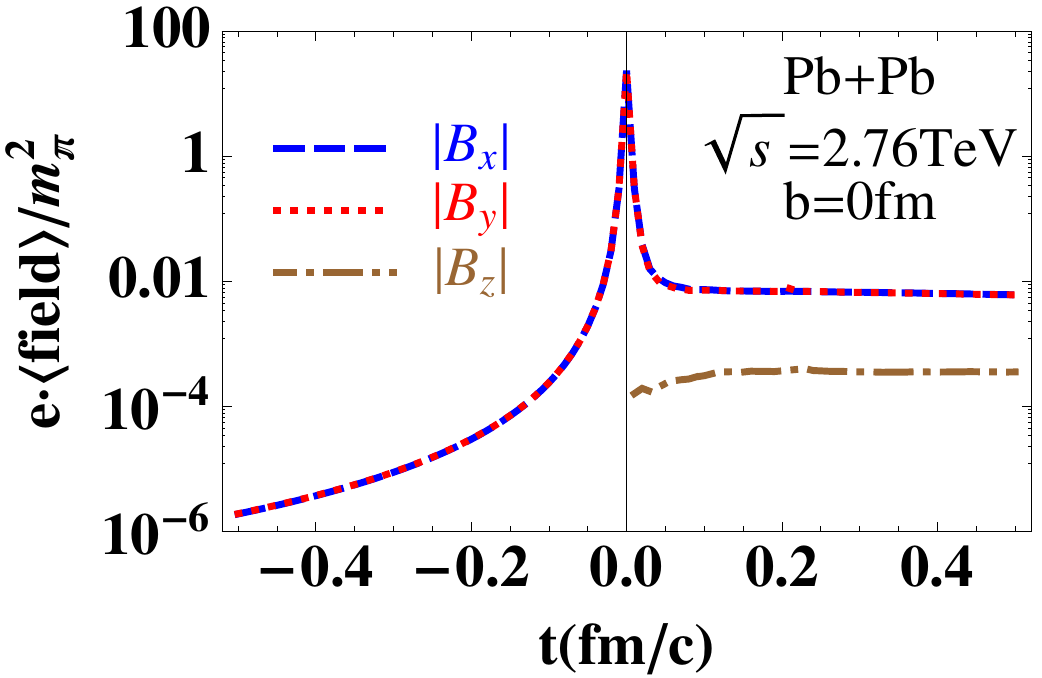}
\includegraphics[width=6.5cm]{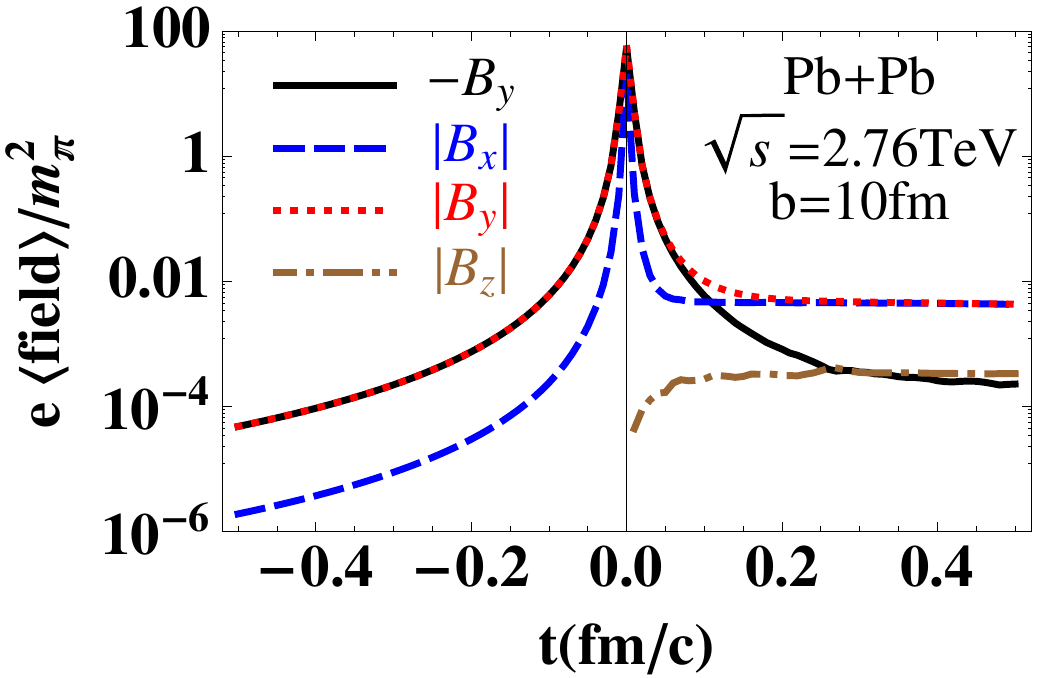}
\includegraphics[width=6.5cm]{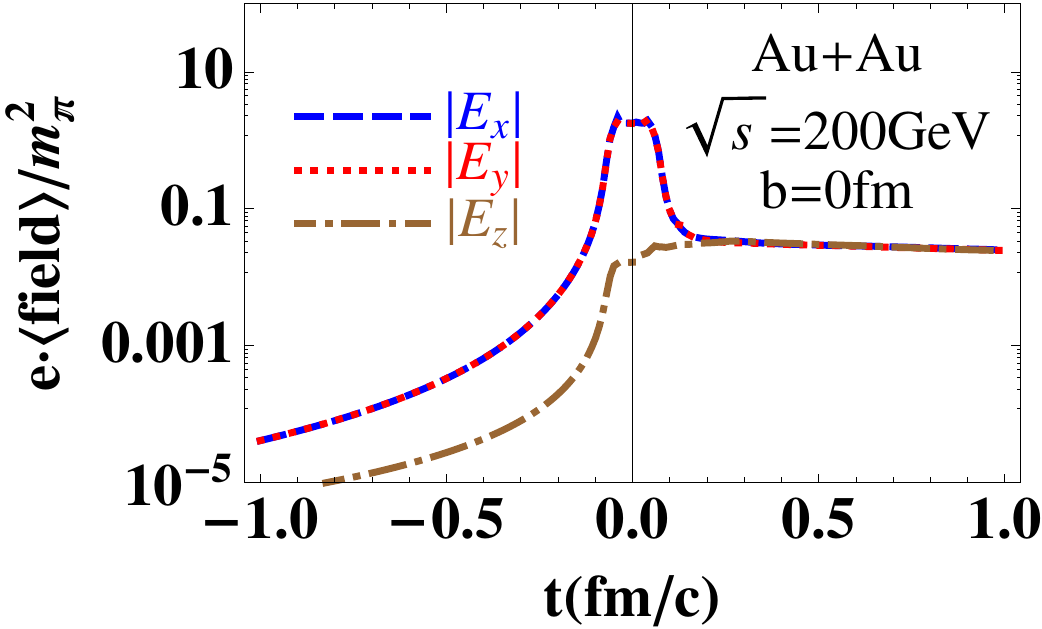}
\includegraphics[width=6.5cm]{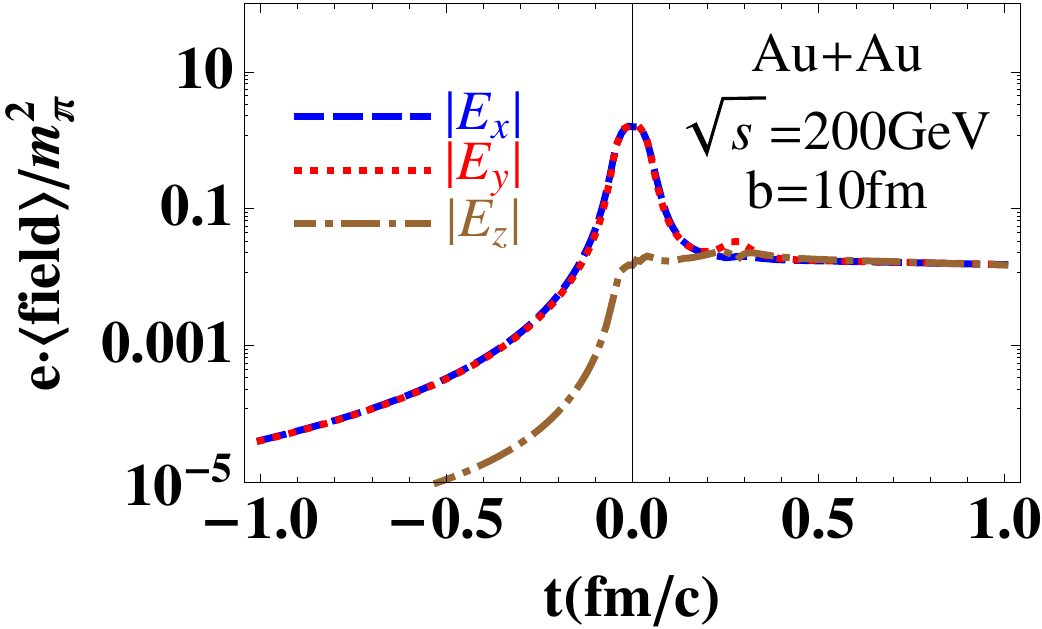}
\includegraphics[width=6.5cm]{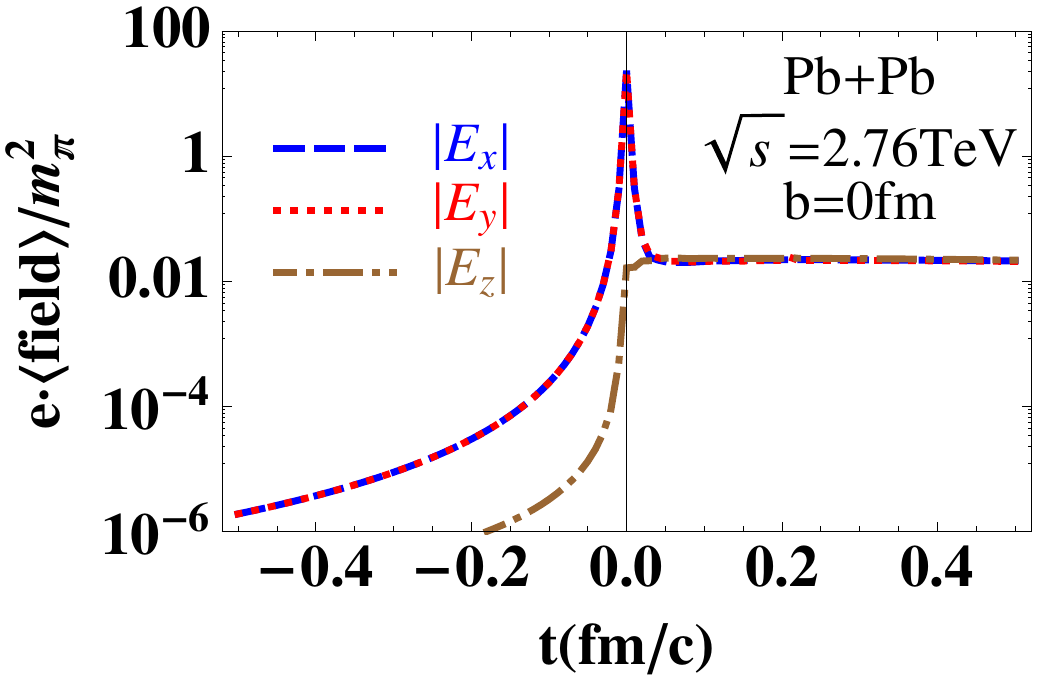}
\includegraphics[width=6.5cm]{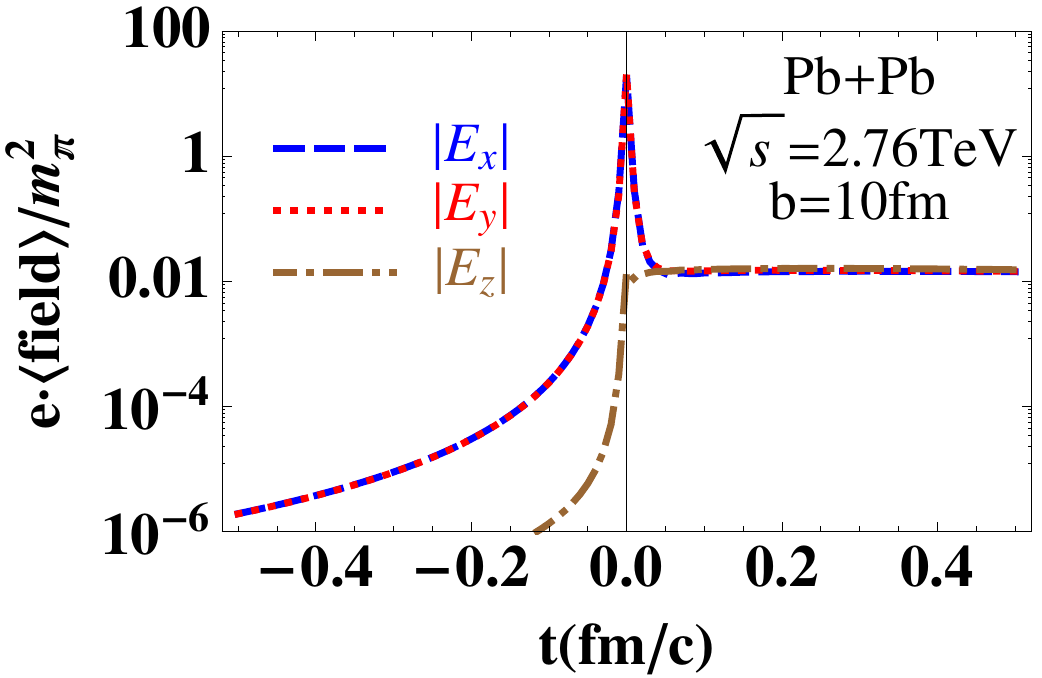}
\caption{(Color online) The time evolution of electromagnetic fields at $\br=0$
with impact parameter $b=0$, and $b=10$ for Au + Au collisions at $\sqrt{s}=200$ GeV and
Pb + Pb collisions at $\sqrt{s}=2.76$ TeV. After collision, the remnants can essentially
slow down the decay of the transverse fields, and enhance the longitudinal fields.}
\label{tdepe}
\end{center}
\end{figure*}
In \fig{tdepe}, we show our results of the early-stage time evolution
of the electromagnetic fields at $\br={\bf0}$ in both
central collisions and off-central collisions with $b=10$ fm, for Au + Au collision at $\sqrt{s}=200$ GeV
and for Pb + Pb collision at $\sqrt{s}=2.76$ TeV.
We take into account
the contributions from charged particles
in spectators, participants, and remnants. Around $t=0$, we checked that the contributions from the
remnants are negligibly small, while
the contributions from participants can be as large as that from spectators. However,
at a latter time when the spectators have already moved far away from the collision region,
the contributions from the remnants become important because they move much slower than the spectators.
These remnants can essentially
slow down the decay of the transverse fields, as seen from \fig{tdepe}.
Another evident effect of the remnants is the substantial enhancements of the longitudinal magnetic
and electric fields caused by the position fluctuation of the remnants, which
have non-zero transverse momenta.
Particularly, although $\lan|B_z|\ran$ is at least one order smaller than
$\lan|B_{x,y}|\ran$ for $t\lesssim1$ fm/c, $\lan|E_z|\ran$ can
evolve to the same amount of $\lan|E_{x,y}|\ran$ in a very short time after collision and
then they decay very slowly.

For central collision, all the fields
are generated due to the position fluctuations of the charged particles.
As seen from \fig{tdepe}, these
fluctuations lead to sizable $\lan|E_x|\ran=\lan|E_y|\ran$ and
$\lan|B_x|\ran=\lan|B_y|\ran$ around $t=0$, but they drop very fast. Note that the fields drop faster
for larger collision energy $\sqrt{s}$.

For off-central collisions, the $y$-component of the magnetic field are much larger than other fields
at $t=0$. But at a latter time when the spectators move far away, the contributions of remnants dominate,
and lead to $\lan|B_y|\ran\approx\lan|B_x|\ran>\lan B_y\ran$.

A common feature of all the fluctuation-caused transverse fields is that they all increase very fast before collision (due to
the fast approaching of the nuclei),
then they drop steeply after $t=0$ (due to the high-speed leaving of the spectators away from the collision center),
and then decay very slowly (due to that contribution from the slowly moving remnants take over that from the spectators).
After the early-stage evolution, the produced QGP may get enough time to respond to the electromagnetic fields, which may
substantially modify the picture of evolution. We discuss this point in the next section.

\section {Response of the QGP to electromagnetic fields}\label{respo}
In the calculations above, we have neglected the electromagnetic response
of the matter produced in the collision (\ie, we assumed the produced matter
is ideally electrically insulating). However, if the produced matter, after a short early-stage evolution,
is in the QGP phase,
the electric conductivity $\s$ is not negligible.
At high temperature, the perturbative QCD
gives that $\s\approx 6T/e^2$~\cite{Arnold:2003zc}, and the lattice calculations give that
$\s\approx 7 C_{\rm EM}T$~\cite{Gupta:2003zh}, or
$\s\approx 0.4 C_{\rm EM}T$~\cite{Aarts:2007wj,Ding:2010ga}, or
$\s\approx (1/3) C_{\rm EM}T$-$C_{\rm EM}T$~\cite{Ding:2010ga,Francis:2011bt}
for temperature of several $T_c$,
where $C_{\rm EM}\equiv \sum_f e_f^2$, $f=u,d,s$, and $e_f$ is the charge of quark with flavor $f$.
Thus it is expected that the QGP can have non-trivial
electromagnetic response. Such electromagnetic response can substantially influence the time
evolution of the electromagnetic fields in the QGP.

To have an estimation of the electromagnetic response of QGP, we
use the following Maxwell's equations,
\begin{eqnarray}
\label{maxwell}
&\displaystyle\nabla\times\bE=-\frac{\pt\bB}{\pt t},&\\
&\displaystyle\frac{1}{\m}\nabla\times\bB=\e\frac{\pt\bE}{\pt t}+\bJ,&
\end{eqnarray}
where $\m$ and $\e$ are the permeability and permittivity of the QGP, respectively, and are assumed as
constants.
$\bJ$ is the electric current determined by the Ohm's law,
\begin{eqnarray}
\label{ohm}
\bJ=\s\lb\bE+\bv\times\bB\rb,
\end{eqnarray}
where $\bv$ is the flow velocity of QGP. Using \eq{ohm}, we can
rewrite the Maxwell's equations as
\begin{eqnarray}
\label{induce1}
&\displaystyle\frac{\pt\bB}{\pt t}=\nabla\times(\bv\times\bB)
+\frac{1}{\s\m}\lb\nabla^2\bB-\m\e\frac{\pt^2\bB}{\pt t^2}\rb,&\\
\label{induce2}
&\displaystyle\frac{\pt\bE}{\pt t}+\frac{\pt\bv}{\pt t}\times\bB=\bv\times(\nabla\times\bE)
+\frac{1}{\s\m}\lb\nabla^2\bE-\m\e\frac{\pt^2\bE}{\pt t^2}\rb,&
\end{eqnarray}
where we have used the Gauss's laws $\nabla\cdot\bB=0$ and $\nabla\cdot\bE=\r=0$ with the assumption that
the net electric charge density of the QGP is zero.
Equation (\ref{induce1}) is the {\it induction equation}, which plays a central role in describing
the dynamo mechanism of stellar magnetic field generation. The first terms on the right-hand sides
of \eqs{induce1}{induce2}
are the convection terms, while the last terms are the diffusion terms. The ratio of these two types of
terms are characterized by the magnetic Reynolds number $R_m$,
\begin{eqnarray}
R_m\equiv LU\s\m,
\end{eqnarray}
where $L$ is the characteristic length or time scale of the QGP, $U$ is the characteristic
velocity of the flow.

Because the theoretical result of $\s$ is quite uncertain, the value of $R_m$ is also uncertain. For
example, by assuming $\m=\e=1$, setting the characteristic length scale $L\sim10$ fm,
and the typical velocity $U\sim 0.5$, we can
estimate $R_m$ at $T=350$ MeV as
$R_m\sim 0.2$ if we use $\s\approx 0.4C_{\rm EM}T$ in Refs.~\cite{Aarts:2007wj,Ding:2010ga},
or $R_m\sim 4$ if we use $\s\approx 7C_{\rm EM}T$ in Ref.~\cite{Gupta:2003zh}, or
$R_m\sim 600$ if we use $\s\approx 6T/e^2$ in Ref.~\cite{Arnold:2003zc}.

If $R_m\ll1$, we can neglect the convection terms in \eq{induce1} and \eq{induce2}. Tuchin studied this
case~\cite{arXiv:1006.3051} (with additional condition $U\ll1$ so that the second-order
time-derivative terms in the diffusion terms are
neglected), and concluded that the magnetic field can be considered as approximately stationary during
the QGP lifetime\footnote{To reach this conclusion, Tuchin used $\s=6T^2/T_c$ to estimate
the magnetic diffusion time $\t=(L/2)^2\s/4$
and found that, for $L=10$ fm and $T=2T_c\approx 400$MeV, $\t\approx 150$ fm is much longer than the lifetime of the QGP. However,
if, for example, $\s\approx 0.4C_{\rm EM}T$ is used,
the magnetic diffusion time is $\t\approx 0.3$ fm, which is much shorter than what Tuchin obtained.}.

If $R_m\gg1$, we can neglect the diffusion terms in \eqs{induce1}{induce2}, \ie,
take the {\it ideally conducting} limit,
\begin{eqnarray}
\label{induce1new}
&\displaystyle\frac{\pt\bB}{\pt t}=\nabla\times(\bv\times\bB),&\\
\label{induce2new}
&\bE=-\bv\times\bB.&
\end{eqnarray}
It is well-known that \eq{induce1new} leads to the {\it frozen-in theorem} for ideally conducting plasma
(\ie, the magnetic lines are frozen in the
plasma elements or more precisely the magnetic flux through a closed loop defined by plasma elements keeps
constant~\cite{Jackson}).

We now use \eqs{induce1new}{induce2new} to estimate how the electromagnetic field evolves in a QGP with
$R_m\gg1$. To this purpose, we have to know the evolution of $\bv$ first. By assuming that the bulk evolution
of the QGP is governed by strong dynamics, we can neglect the influence
of the electromagnetic field on the evolution of the velocity $\bv$.
We assume the Bjorken picture for the longitudinal expansion,
\begin{eqnarray}
\label{vz}
v_z=\frac{z}{t}.
\end{eqnarray}
Because the early transverse expansion is slow, following Ref.~\cite{Ollitrault:2007du},
we adopt a linearized ideal hydrodynamic equation to describe the transverse flow velocity $\bv_\perp$,
\begin{eqnarray}
\label{hydro}
\frac{\pt}{\pt t}\bv_\perp=-\frac{1}{\ve+P}\nabla_\perp P=-c_s^2\nabla_\perp \ln {\mathfrak s},
\end{eqnarray}
where we used $\ve+P=T {\mathfrak s}$, $\mathfrak{s}$ is the entropy density, and $c_s=\sqrt{\pt P/\pt\ve}$ is the speed of sound.
For simplicity, we choose an initial Gaussian transverse entropy density profile as in~\cite{Ollitrault:2007du},
\begin{eqnarray}
{\mathfrak s}(x,y)={\mathfrak s}_0\exp{\lb-\frac{x^2}{2a^2_x}-\frac{y^2}{2a^2_y}\rb},
\end{eqnarray}
where $a_{x,y}$ are the root-mean-square widths of the transverse distribution. They are of order of the nuclei radii if the
impact parameter is not large. For example, for Au + Au collisions at RHIC,
$a_x\sim a_y\sim3$ fm for $b=0$, and $a_x\sim2$ fm, $a_y\sim3$ fm for $b=10$ fm.
One can then easily solve \eq{hydro} and obtain,
\begin{eqnarray}
\label{vx}
v_x&=&\frac{c_s^2}{a^2_x}xt,\\
\label{vy}
v_y&=&\frac{c_s^2}{a^2_y}yt.
\end{eqnarray}
Substituting the velocity fields above into \eq{induce1new},
we obtain a linear differential equation for $\bB(t)$. For a given initial
condition $\bB^0(\br)=\bB(t=t_0,\br)$ where $t_0$ is
the formation time of the QGP, it can be solved analytically,
\begin{eqnarray}
\label{evobx}
B_x(t,x,y,z)&=&\frac{t_0}{t}e^{-\frac{c_s^2}{2a_y^2}(t^2-t_0^2)}B_x^0\lb x e^{-\frac{c_s^2}{2a_x^2}(t^2-t_0^2)},
y e^{-\frac{c_s^2}{2a_y^2}(t^2-t_0^2)},z\frac{t_0}{t}\rb,\\
\label{evoby}
B_y(t,x,y,z)&=&\frac{t_0}{t}e^{-\frac{c_s^2}{2a_x^2}(t^2-t_0^2)}B_y^0\lb x e^{-\frac{c_s^2}{2a_x^2}(t^2-t_0^2)},
y e^{-\frac{c_s^2}{2a_y^2}(t^2-t_0^2)},z\frac{t_0}{t}\rb.
\end{eqnarray}
Because $B_z$ is always much smaller than $B_x$ and $B_y$, we are not interested in it.
The electric fields can be obtained from \eq{induce2new} once we have $\bB(t,\br)$.

To reveal the physical content in \eqs{evobx}{evoby}, we notice that, by
integrating \eq{vz}, \eq{vx}, and \eq{vy}, a fluid cell located at $(x_0,y_0,z_0)$ at time $t_0$ will flow to the coordinate
$(x,y,z)$ at time $t$ with
\begin{eqnarray}
x&=&x_0 \exp{\ls\frac{c_s^2}{2a_x^2}(t^2-t_0^2)\rs},\\
y&=&y_0 \exp{\ls\frac{c_s^2}{2a_y^2}(t^2-t_0^2)\rs},\\
z&=&z_0\frac{t}{t_0}.
\end{eqnarray}\
Thus,we can rewrite \eqs{evobx}{evoby} as
\begin{eqnarray}
\label{evobx0}
B_x(t,x,y,z)&=&\frac{t_0}{t}e^{-\frac{c_s^2}{2a_y^2}(t^2-t_0^2)}B_x\lb t_0,x_0,y_0,z_0\rb,\\
\label{evoby0}
B_y(t,x,y,z)&=&\frac{t_0}{t}e^{-\frac{c_s^2}{2a_x^2}(t^2-t_0^2)}B_y\lb t_0,x_0,y_0,z_0\rb.
\end{eqnarray}
As the areas of the
cross sections of the QGP expand according to
$t\,\exp{\lb\frac{c_s^2}{2a_y^2}t^2\rb}$ in the $y$-$z$ plane and $t\,\exp{\lb\frac{c_s^2}{2a_x^2}t^2\rb}$ in the $x$-$z$ plane,
\eqs{evobx0}{evoby0} mean that the magnetic line flows with the fluid cell and is diluted due to the expansion of the QGP.
These are just the manifestations of the frozen-in theorem,
We also note that \eqs{evobx}{evoby} can be written in explicit scaling forms,
\begin{eqnarray}
\label{evobx1}
t\,e^{\frac{c_s^2}{2a_y^2}t^2}B_x\lb t,x e^{\frac{c_s^2}{2a_x^2}t^2},
y e^{\frac{c_s^2}{2a_y^2}t^2},z\,t\rb&=&t_0\,e^{\frac{c_s^2}{2a_y^2}t_0^2}B_x\lb t_0,
x e^{\frac{c_s^2}{2a_x^2}t_0^2},
y e^{\frac{c_s^2}{2a_y^2}t_0^2},z\,t_0\rb,\\
\label{evoby1}
t\,e^{\frac{c_s^2}{2a_x^2}t^2}B_y\lb t,x e^{\frac{c_s^2}{2a_x^2}t^2},
y e^{\frac{c_s^2}{2a_y^2}t^2},z\,t\rb&=&t_0\,e^{\frac{c_s^2}{2a_x^2}t_0^2}B_y\lb t_0,
x e^{\frac{c_s^2}{2a_x^2}t_0^2},
y e^{\frac{c_s^2}{2a_y^2}t_0^2},z\,t_0\rb.
\end{eqnarray}

From \eqs{evobx}{evoby}, we see that the evolution
of $\bB$ is strongly influenced by its initial spatial distribution. However,
the time evolution of the magnetic fields at the center of
the collision region, $\br={\bf 0}$, takes very simple forms,
\begin{eqnarray}
\label{bxevo}
B_x(t, {\bf 0})&=&\frac{t_0}{t}e^{-\frac{c_s^2}{2a_y^2}(t^2-t_0^2)}B_x^0({\bf 0}),\\
\label{byevo}
B_y(t, {\bf 0})&=&\frac{t_0}{t}e^{-\frac{c_s^2}{2a_x^2}(t^2-t_0^2)}B_y^0({\bf 0}).
\end{eqnarray}
Setting $a_x\sim a_y\sim3$ fm and $c_s^2\sim1/3$, we see from \eqs{bxevo}{byevo}
that for $t\lesssim5$ fm the magnetic fields decay inversely proportional to time.

\section {Summary and Discussions}\label{discu}
In summary, we have utilized the HIJING model to investigate the generation and evolution of the
electromagnetic fields in heavy-ion collisions. The cases of Au + Au collision at $\sqrt{s}=200$ GeV
and Pb + Pb collision at $\sqrt{s}=2.76$ TeV are considered in detail.
Although after averaging over many events only the component $B_y$ remains, the event-by-event
fluctuation of the positions of charged particles can induce
components $B_x, E_x, E_y $ as large as $B_y$. They can reach
the order of several $m_\p^2/e$.
The spatial structure of the electromagnetic field is studied and a very inhomogeneous distribution is found.
We study also the time evolution of the fields including the early-stage and the QGP-stage evolutions.
We find that the remnants can give substantial contribution to the fields during the early-stage evolutions.
The non-trivial electromagnetic response of the QGP, which is sensitive to the electric conductivity, gives
non-trivial time dependence of the fields in it, see Sec.~\ref{respo}.
We check both in numerical calculation (\fig{cedepe}) and analytical derivations (\eqs{et0}{bt0})
that the electric and magnetic fields at $t=0$ have approximately linear dependence on the collision energy $\sqrt{s}$.

The strong event-by-event fluctuation of
the electromagnetic field may lead to important implications for observables which are sensitive to
the electromagnetic field. We point out two examples here.

(1) From \fig{bdepe} we see that although the electric fields $E_x$ and $E_y$ at $\br={\bf 0}$ can
be very strong, they are roughly equal. Then one expects that
the strong electric fields should not have a significant contribution for the correlation observable
$\lan\cos(\f_1+\f_2-2\Psi_{RP})\ran$ which is sensitive to the chiral magnetic effect~\cite{Voloshin:2004vk}, where $\f_{1,2}$ are
the azimuthal angles of the final-state charged particles, and $\Psi_{RP}$ is the azimuthal angle of the reaction plane.
On the other hand, from \fig{sdist} we see that in the overlapping region for peripheral collisions,
the electric field perpendicular to the reaction plane
has larger gradient than that parallel to the reaction plane. Thus, a strong, out-of-plane
electric field develops away from the origin $\br={\bf0}$ in the overlapping region. Note that the direction of
this electric field
points outside of the reaction plane.
If the electric conductivity of the matter
produced in the collision is large, this out-of-plane electric field can drive positive (negative) charges to
move outward (toward)
the reaction plane, and thus induce an electric quadrupole moment. Such an electric quadrupole moment,
as argued in Ref.~\cite{arXiv:1103.1307}, can lead to an elliptic flow imbalance between $\p^+$ and $\p^-$. Thus, it will be interesting to
study how strong the electric quadrupole moment can be induced by this out-of-plane electric field. Note that such electric quardrupole configuration
does not contribute to the correlation $\lan\cos(\f_1+\f_2-2\Psi_{RP})\ran$.

(2) It is known that quarks produced in off-central heavy-ion collision can be possibly polarized due to
the spin-orbital coupling of QCD~\cite{Liang:2004ph,Gao:2007bc,Huang:2011ru}.
The strong magnetic field can cause significant polarization of quarks as well through the interaction
between the quark magnetic moment and the magnetic field. As estimated
by Tuchin~\cite{arXiv:1006.3051}, a magnetic field of order $m_\p^2/e$ can almost immediately polarize
light quarks. Such polarization, contrary to the polarization due to spin-orbital coupling,
will depend on the charges of quarks, and build a spin-charge correlation for quarks, \ie, the
positively charged quarks are polarized along the magnetic field while the negatively charged quarks are
polarized opposite to the magnetic field.
If we expect that the strong interaction
in the QGP and the hadronization processes do not wash out this spin-charge correlations, we should observe
similar spin-charge correlation for final-state charged hadrons.

{\bf Acknowledgments:} We thank A. Bzdak, V. Skokov, H. Warringa, and Z. Xu for helpful discussions and comments.
This work is supported by the Helmholtz International Center for FAIR within the
framework of the LOEWE program (Landesoffensive zur Entwicklung
Wissenschaftlich- \"Okonomischer Exzellenz) launched by the State of Hesse.
The calculations are partly performed at the Center for Scientific Computing of
J. W. Goethe University. Some of the figures are plotted using LevelScheme
toolkit\cite{Caprio:2005dm} for Mathematica.

\end{document}